# Statistical Learning in Preclinical Drug Proarrhythmic Assessment

Nan Miles Xi [a,*], Yu-Yi Hsu [b], Qianyu Dang [b], and Dalong Patrick Huang [b,*]

[a] *Department of Mathematics and Statistics, Loyola University Chicago, Chicago, IL 60660, USA*

[b] *Office of Biostatistics, Office of Translational Science, Center for Drug Evaluation and Research, US Food and Drug Administration, Silver Spring, MD 20993, USA*

[*] Correspondence: Nan Miles Xi (mxi1@luc.edu)

Dalong Patrick Huang (Dalong.Huang@fda.hhs.gov)

# Statistical Learning in Preclinical Drug Proarrhythmic Assessment

**ABSTRACT**

Torsades de pointes (TdP) is an irregular heart rhythm characterized by faster beat rates and potentially could lead to sudden cardiac death. Much effort has been invested in understanding the drug-induced TdP in preclinical studies. However, a comprehensive statistical learning framework that can accurately predict the drug-induced TdP risk from preclinical data is still lacking. We proposed ordinal logistic regression and ordinal random forest models to predict low-, intermediate-, and high-risk drugs based on datasets generated from two experimental protocols. Leave-one-drug-out cross-validation, stratified bootstrap, and permutation predictor importance were applied to estimate and interpret the model performance under uncertainty. The potential outlier drugs identified by our models are consistent with their descriptions in the literature. Our method is accurate, interpretable, and thus useable as supplemental evidence in the drug safety assessment.

**Introduction**

Torsades de pointes (TdP) is a rare but potentially fatal ventricular arrhythmia largely caused by electrolyte imbalance and cardiomyopathies after drug treatment.[1] TdP typically occurs along with the prolongation of the interval between the start of the Q-wave and the end of the T-wave (QT) in the electrocardiogram (ECG) and results in polymorphic ventricular tachycardia.[2] The identification of TdP is a crucial step in the assessment of safety before a drug reaches the market.[3] The International Council on Harmonisation (ICH) proposed two regulatory guidelines, ICH S7B and ICH E14, to assess the drug-induced TdP risk in preclinical and clinical studies, respectively.[4,5] The ICH S7B recommends using an ionic channel, human ether-a-go-go-related gene (hERG) current, as the proxy to evaluate drug-induced TdP risk *in vitro*. Drugs blocking the hERG current are considered candidates with high TdP risk. The ICH E14 outlines a framework to examine the drug's capacity to prolong QT intervals *in vitro*. High TdP risk is assigned to drugs that induce the prolonged QT intervals in the ECG. Both guidelines focus on a single indicator of TdP risk and show high sensitivity (true positive rate) in the identification of drugs with high TdP risk. However, the overemphasis of hERG block or QT prolongation alone leads to conservative assessment and low specificity (true negative rate), potentially eliminating safe, promising drugs from the market.[6]

Previous studies revealed the factors beyond the hERG block in the assessment of drug-induced TdP risk. For example, inhibiting inward sodium or calcium currents offsets the effects of hERG block, and drugs with such mechanisms show low TdP risk.[7] Realizing the limitation of ICH S7B and ICH E14, the regulatory agency, pharmaceutical industry, and academia proposed several new paradigms to incorporate this multichannel interaction for better prediction of drug-induced TdP risk in preclinical studies. The Comprehensive *In Vitro* Proarrhythmia Assay (CiPA), initiated by the US Food and Drug Administration (FDA), is an essential public-private collaboration among these attempts.[8] The CiPA initiative relies on the drug effects on human ventricular electrical activity measured by cellular electrophysiology (EP) of human-induced pluripotent stem cell-derived cardiomyocytes (hiPSC-CMs).[9] It incorporates multiple ion channels beyond hERG and *in silico* modeling to predict drug-induced TdP risk. Twenty-eight drugs with clearly understood TdP risks in the market were selected as the benchmark set for the validation of the CiPA initiative. Additionally, the rabbit ventricular wedge assay (RVWA), an established *in vitro* paradigm for detecting drug-induced QT prolongation and arrhythmia,[10,11] has been adapted for the assessment of drug-induced TdP risk.[12,13] RVWA relies on the drug effects on the rabbit left ventricular wedge measured by pseudo-ECG. A previous study applied RVWA to 28 CiPA drugs and exhibited promising assessment results using a normalized TdP score.[14]

Built on the development of *in vitro* paradigms, *in silico* models for predicting drug-induced TdP risk have been promoted in numerous studies.[15–18] The prediction is essentially a classification problem fit into the statistical learning framework — given the input data of one drug from the *in vitro* study, the model outputs the risk of that drug. Therefore, the principles of statistical learning should be accommodated into the design of such models.[19] First, the model is expected to output both classifying and ranking measurements of drug-induced TdP risk. Second, the uncertainty of such measurements is required to incorporate the variation in the experiment and modeling. Third, the model must provide an unbiased estimate of prediction performance on drugs outside the available data. Fourth, the statistical learning model itself should be able to capture both linear and nonlinear relationships in the data. Finally, the model prediction can be easily interpreted to offer mechanistic insights into the drug-induced TdP risk. Current *in silico* models mainly use score-based decision rules or simple linear regression methods. A statistical learning method satisfying the five principles is lacking.

In this study, we proposed two statistical learning models, ordinal logistic regression and ordinal random forest, to accurately predict drug-induced TdP risk on datasets generated under CiPA and RVWA paradigms. Our predictive models utilized the ordinal information in low-, intermediate-, and high-risk levels instead of treating them as independent categories. The unbiased model performance on new drugs was estimated by leave-one-drug-out cross-validation (LODO-CV). The uncertainty of model performance was further quantified by stratified bootstrap. We identified the potential outlier drugs using the asymptotic prediction accuracy obtained from stratified bootstrap. Sensitivity analysis was then conducted to investigate the impact of potential outlier drugs on the model performance. Specifically, we compared the model prediction accuracy before and after removing potential outlier drugs from the dataset. To further validate and improve model prediction, we conducted control analysis, a common practice in *in vitro* studies, by selecting one control drug with mechanistically understood TdP risk. Finally, we examined the model interpretability through the analysis of normalized permutation predictor importance.

The proposed modeling strategies were evaluated on datasets from two recently published studies — the stem cell dataset under the CiPA paradigm and the wedge dataset under the RVWA paradigm.[6,14] Both datasets contain the same 28 CiPA drugs with known low, intermediate, or high TdP risks (Supplementary Table S1) and multiple observations for each drug. The ordinal logistic regression and ordinal random forest exhibited similar prediction performance on the stem cell dataset in terms of four measurements. On the other hand, ordinal random forest consistently outperformed ordinal logistic regression on the wedge dataset measured by the same four measurements. The identified potential outlier drugs were consistent with the abnormal observations reported in the original studies. The model performance significantly improved after removing potential outlier drugs from the original datasets. We also found that the two models performed better conditioned on the correct prediction of the control drug. The model specification was further validated by the top predictors in terms of normalized permutation predictor importance. In summary, our work is the first attempt to construct multivariate statistical learning models that can accurately predict the drug-induced TdP risk from *in vitro* data. It satisfies the principles of statistical learning, highlighted by its comprehensive uncertainty measurements and strong interpretability. The proposed modeling and evaluation process can be extended easily to new datasets generated by other experimental protocols. The result of model prediction will serve as supplemental evidence in the drug safety assessment.

**Datasets**

*Stem cell dataset*

The first dataset was generated from a blinded *in vitro* study that aimed to assess the drug-induced TdP risk on hiPSC-CMs.[6] The experiments in this study contain 28 drugs selected by the CiPA initiative with known low, intermediate, and high TdP risks. Electrophysiological responses of hiPSC-CMs to 28 drugs at certain concentrations were recorded, resulting in seven predictors in the dataset (Supplementary Table S2). There are two commercial human cardiomyocyte lines, five EP platforms, and 10 sites involved in the experiments. For each drug, 15 observations were recorded based on the combinations of sites, cell lines, and electrophysiological devices. In total, there are 420 observations in this dataset (15 observations × 28 drugs), and we refer to it as the stem cell dataset in the following text.

*Wedge Dataset*

The second dataset was generated from a blinded *in vitro* study of RVWA for the assessment of drug-induced proarrhythmia.[14] The same 28 CiPA drugs were applied to RVWA, and 15 predictors were obtained from the corresponding electrophysiological responses (Supplementary Table S3). For each drug, four observations were collected in one laboratory under the same biological model and electrophysiological device. There are 112 observations in this dataset (4 observations × 28 drugs), and we refer to it as the wedge dataset in the following text.

There are four major differences between these two datasets. First, the stem cell dataset adopts hiPSC-CMs as the biological model, while the wedge dataset uses RVWA. Second, the stem cell dataset was generated from a multisite experiment. On the other hand, the wedge dataset was collected in a single laboratory. Third, the sample size and the number of replications for each drug are roughly four times larger in the stem cell dataset than the wedge dataset. Forth, the stem cell dataset contains seven predictors selected by domain experts, while the wedge dataset includes 15 different predictors.

**Methods**

*Missing value imputation*

The stem cell and wedge datasets contain missing values due to the incomplete recording of raw electrophysiological signals in the experiments. To solve this issue, we utilized the association among predictors to impute missing values in both datasets. First, we trained a bagged tree model,[20] an imputation method that has shown better performance over others in previous studies,[21] by treating predictors with missing values as dependent variables and other predictors as independent variables. Second, we used the pretrained bagged tree model to predict the missing values through the complete predictors and observations. It is worth noting that we did not use the TdP risk information of 28 drugs in the imputation process. Therefore, the information of the prediction target, drug-induced TdP risk, did not leak to the predictors through imputation, thereby avoiding the overfitting in the model prediction. In the imputation process, we include all seven predictors in the stem cell dataset and all 15 predictors in the wedge dataset.

*Model evaluation strategy*

The model evaluation aims to estimate the prediction accuracy of new drugs not contained in the dataset that helped train the model. A valid evaluation strategy should generate an unbiased estimate of the model performance irrespective of the model specification. To achieve this goal, we designed an LODO-CV strategy (Figure 1).[19] For each dataset in this study, we first trained a predictive model on the observations of 27 drugs and predicted the risk of observations in the one left-out drug. Then we repeated the same process by treating each drug as the left-out drug and predicting its observations by models trained on the observations of the other 27 drugs. Finally, we combined the prediction result of each drug and calculated the model performance measurement.

We take the calculation of three-category prediction accuracy as an example to illustrate the proposed LODO-CV. Let $N$ be the number of drugs in the dataset ($N$ = 28 in both datasets) and $J$ be the number of observations per drug ($J$ = 15 in the stem cell dataset; $J$ = 4 in the wedge dataset). Denote $\hat{f}^{-k}$ as the predictive model trained on the dataset *without* the observations of drug $k$. Let $(x_j^k, y_j^k)$ be the $j$th observation of drug $k$, where $x$ and $y$ refer to the predictor vector and risk category, respectively. Then the three-category prediction accuracy under LODO-CV, $acc_{LODO-CV}$, is calculated as:

$$acc_{LODO-CV} = \frac{1}{N}\sum_{k=1}^{N}\frac{1}{J}\sum_{j=1}^{J}I(y_j^k = \hat{f}^{-k}(x_j^k))$$

where $I(x)$ is the indicator function.

The prediction performance estimated by LODO-CV has three advantages. First, unlike previous studies that only used a small proportion of drugs in the dataset to train the model,[18] LODO-CV maximizes the size of the training set, i.e., 27 drugs, and thus reduces the power loss in the validation. Second, it eliminates the model overfitting because there are no observations of the same drug in both training and test sets. Third, the prediction of left-out drugs imitates the scenario in which the model predicts a new drug beyond the existing data. It should be emphasized that LODO-CV is to provide an unbiased estimate of model performance. In the real-world application, we train the model once on all available data (e.g., 28 drugs in our study) and predict observations/drugs generated from new experiments.

*Utilization of ordinal information*

The prediction of drug-induced TdP risk is essentially a three-category classification problem. Given one vector of predictors (one observation) collected from experiments, the predictive model will output one of the three risk categories. Different from vanilla classification tasks,[19] the three risk categories are not independent — ordinal information exists in the risk categories, i.e., low risk < intermediate risk < high risk. Previous benchmark studies have shown that the appropriate utilization of ordinal information can improve the model prediction power.[22,23] In the statistics and machine learning literature, there are several methods specifically designed for ordinal classification.[64] Among them, the regression-based method and binary decomposition method are commonly used in practice. The regression-based method generalizes the logistic regression from the binary response to the response with $k$ ordinal levels.[62] For each level $j$, it models the log odd ratio between "response is less or equal than $j$" and "response is greater than $j$" as a linear combination of predictors. It also assumes the same parameters in different linear combinations for each level $j$, only allowing the intercept to vary. This assumption is often named "proportional odds assumption" in the literature.[62] In the following text, we will refer to the regression-based method as "classical ordinal logistic regression".

The binary decomposition method divides the ordinal response into multiple binary responses and fits the same or different types of models to solve those binary classification tasks.[24] With $k$ ordinal levels, it splits one ordinal classification into $k$-1 binary classification tasks, that is, {1} vs. {2, … k}, {1, 2} vs. {3, …, k}, …, {1, …, $k$-1} vs. {$k$}. After simple algebra, we can obtain the probabilities of all $k$ levels using the outputs of $k$-1 binary classifiers. For example, $p(2) = p(1, 2)$

– $p(1)$, $p(3) = p(1, 2, 3) – p(1, 2)$, and so on. The binary decomposition method does not require proportional odds assumption and can deploy any models in the *k*-1 binary classification tasks. We will adopt the binary decomposition method in our study because of its flexibility and assumption-free characteristics. The following text will refer to it as "ordinal model", where the specific name of binary classifiers would substitute "model".

In the setting of drug TdP risk, we utilized the ordinal information by transforming the three-category classification problem into two binary classification problems. A schematic diagram of this algorithm is described in Figure 2. First, we trained a binary classifier $\hat{f}_1$ which differentiated observations between low risk and intermediate-or-high risk. Similarly, we trained another binary classifier $\hat{f}_2$ which differentiated observations between high risk and low-or-intermediate risk. Second, for any observation $x$ in the test set (i.e., the left-out drug), we predicted its probability of low risk $p_x(L)$, intermediate-or-high risk $p_x(MH)$, high risk $p_x(H)$, and low-or-intermediate risk $p_x(LM)$ as:

$$p_x(L) = \hat{f}_1(x),$$

$$p_x(MH) = 1 - p_x(L),$$

$$p_x(H) = \hat{f}_2(x),$$

$$p_x(LM) = 1 - p_x(H).$$

Then the probability of intermediate risk $p_x(M)$ was calculated as

$$p_x(M) = p_x(MH) - p_x(H).$$

Finally, the risk of *observation* $x$, $\widehat{risk}(x)$, was predicted as

$$\widehat{risk}(x) = argmax_r\{p_x(r)\}$$

where $r \in \{L, M, H\}$. To predict the *drug* risk, we averaged the risk probability of observations in each drug and selected the risk category with the highest probability. Let $J$ be the number of observations per drug and $\{x_j^k\}$ be the observations of drug $k$, where $j = 1, 2, ..., J$. Given the observations' risk probabilities $\{p_{x_j^k}(L), p_{x_j^k}(M), p_{x_j^k}(H)\}$, the drug $k's$ probabilities of low risk $p_k(L)$, intermediate risk $p_k(M)$, and high risk $p_k(H)$ were calculated as:

$$p_k(L) = \frac{1}{J}\sum_{j=1}^{J} p_{x_j^k}(L)$$

$$p_k(M) = \frac{1}{J}\sum_{j=1}^{J} p_{x_j^k}(M)$$

$$p_k(H) = \frac{1}{J}\sum_{j=1}^{J} p_{x_j^k}(H)$$

Then the risk of drug $k$, $\widehat{risk}(k)$, was predicted as:

$$\widehat{risk}(k) = argmax_r\{p_k(r)\}$$

where $r \in \{L, M, H\}$. Note that this framework can accommodate any form of binary classifiers if they can output the probability of each risk category.

*Ordinal logistic regression model*

The first model we built is ordinal logistic regression, in which logistic regressions were substituted for the two binary classifiers in the proposed ordinal framework. Formally, for any observation $x$, the two binary classifiers $\hat{f}_1$ and $\hat{f}_2$ in the ordinal framework are defined as:

$$\hat{f}_1: \log \frac{p_x(L)}{1-p_x(L)} = X\hat{\beta}_1$$

$$\hat{f}_2: \log \frac{p_x(H)}{1-p_x(H)} = X\hat{\beta}_2$$

where $X$ is the predictor vector of observation $x$ and $\hat{\beta}_1$ and $\hat{\beta}_2$ are two model parameter vectors (including interceptions). $X\hat{\beta}_1$ and $X\hat{\beta}_2$ are inner products between predictor vectors and parameter vectors. We estimated $\hat{f}_1$ and $\hat{f}_2$ using maximum likelihood estimation.[61] After obtaining $\hat{f}_1$ and $\hat{f}_2$, the prediction of risk categories for each observation and drug were calculated as described in the previous section.

*Ordinal random forest model*

The second model we built is ordinal random forest, in which random forests were substituted for the two binary classifiers in the proposed ordinal framework. Random forest is the ensemble of multiple decision trees and can capture the nonlinear relationship in the dataset.[25] A decision tree $T$ is a predictive model that assigns each observation to a certain category based on split rules defined on the predictor space. Formally, suppose that there are $P$ predictors $X_1, X_2, \ldots, X_P$ in the dataset, and we split the predictor space into two regions, $R_1$ and $R_2$, according to predictor $X_t$ and threshold $s$:

$$R_1(x, t, s) = \{x | X_t \leq s\}$$

$$R_2(x, t, s) = \{x | X_t > s\}$$

where $x$ denotes observation. Then for any region $R_m$ with $N_m$ observations, let $\hat{p}_{mr}$ be the proportion of category $r$ in region $R_m$:

$$\hat{p}_{mr} = \frac{1}{N_m} \sum_{x_i \in R_m} I(y_i = r)$$

where $x_i$ and $y_i$ are the predictor vector and risk category of observation $i$, respectively. $I(x)$ is the indicator function. The risk of any observation $x$ in region $R_m$ is predicted as:

$$\widehat{risk}(x) = argmax_r \hat{p}_{mr},$$

where $r \in \{L, M, H\}$. In each split generating regions $R_1$ and $R_2$, we seek the predictor $X_t$ and threshold $s$ by solving the following optimization task:

$$min_{t,s}[\sum_{x_i \in R_1(t,s)} L(\widehat{risk}(x_i), y_i) + \sum_{x_l \in R_2(t,s)} L(\widehat{risk}(x_l), y_l)]$$

where $L(x, y)$ is the loss function that can be misclassification error, Gini index, or cross-entropy.[19] The splitting process continues until it satisfies certain stopping rules, usually the maximum number of splits (tree depth) or the minimum number of observations per region (leaf size). In our implementation, we set the leaf size to 1, the default value in the R package randomForest.[26]

To build a random forest from multiple decision trees, we generated $B$ ($B = 500$ in our implementation) bootstrap samples from the training set. We applied the previous splitting rule to build one decision tree for each bootstrap sample. Instead of searching all $P$ predictors, we randomly selected $\sqrt{P}$ predictors in each split to reduce the correlation among different decision trees. The final random forest $\{T_b\}_1^B$ contains $B$ decision trees, and the prediction of risk is the majority vote of all decision trees. As random forest is a combination of decision trees with low correlations, it reduces the high prediction variance of a single decision tree and thus exhibits high prediction accuracy in real-world applications.[27]

*Model performance measurements*

We quantified the model performance by four measurements: three-category prediction accuracy, two areas under the receiver operating characteristics curve (AUROCs) (high risk vs. intermediate-or-low risk and high-or-intermediate risk vs. low risk), and concordance index. All measurements were calculated under the LODO-CV measured by both observations and drugs.

The three-category prediction accuracy is to measure the model's capacity of classifying low-, intermediate-, and high-risk observations and drugs. Suppose that there are $N$ observations and $J$ drugs in the dataset. Let $\hat{y}_{i\_obs}$ and $y_{i\_obs}$ be the predicted and true risks for observation $i$, and $\hat{y}_{j\_drug}$ and $y_{i\_drug}$ be the predicted and true risks for drug $j$. The three-category prediction accuracy by observations $acc_{obs}$ and by drug $acc_{drug}$ were calculated as:

$$acc_{obs} = \frac{1}{N}\sum_{i=1}^{N} I(\hat{y}_{i\_obs} = y_{i\_obs})$$

$$acc_{drug} = \frac{1}{J}\sum_{j=1}^{J} I(\hat{y}_{j\_drug} = y_{j\_drug})$$

where $I(x)$ is the indicator function.

The AUROC of high risk vs. intermediate-or-low risk and AUROC of high-or-intermediate risk vs. low risk measure the model's diagnostic ability as two binary classifiers. For each observation and drug, we first calculated its probabilities of low risk $p(L)$, intermediate-or-high risk $p(MH)$, high risk $p(H)$, and low-or-intermediate risk $p(LM)$. Then we plotted the true positive rate (TPR) against the false positive rate (FPR) at various thresholds and calculated the area under that curve. The calculation of two AUROCs was implemented by R package PRROC.[28]

The concordance index evaluates the quality of ranking in the ordinal prediction task. It was calculated as the proportion of concordant risk pairs in all predicted risk pairs by observations or drugs.[29] We calculated the concordance index by using R package reticulate[30] to call Python library lifelines.[31]

**Results**

*Overall model performance*

We used the R programming language[32] to implement the proposed modeling and validation strategies. Table 1 and Supplementary Table S4 summarize the prediction result of ordinal logistic regression and ordinal random forest. On the stem cell dataset, if measured by observations, the ordinal logistic regression has higher AUROC of high risk vs. intermediate-or-low risk (0.817) and concordance index (0.764). On the other hand, the ordinal random forest shows advantages in terms of the three-category prediction accuracy (0.603) and AUROC of high-or-intermediate risk vs. low risk (0.830) (Supplementary Table S4). If measured by drugs, the ordinal logistic regression has higher three-category prediction accuracy (0.643) and concordance index (0.811), while the ordinal random forest shows advantages in terms of two AUROCs (0.900 and 0.889) (Table 1). Overall, the two models exhibit mixed performance on the stem cell dataset. No model consistently outperforms the other, and the gaps of different measurements between the two models range from minor to moderate. The AUROCs and concordance indices of the two models are between 0.750 and 0.850, indicating a good capacity of binary classification and order assignment. On the more difficult three-category classification task, the best prediction accuracy obtained from the same model drops to ~0.600. Since the ordinal random forest can capture the nonlinear relationship,[25] the similar performance between the two models implies the ignorable nonlinearity in the stem cell dataset.

On the wedge dataset, ordinal random forest consistently outperforms ordinal logistic regression in terms of all four measurements measured by observations (Supplementary Table S4) and drugs (Table 1). The highest AUROCs and concordance indices are over 0.900, indicating excellent binary classification and order assignment. On the more difficult three-category classification task, the best prediction accuracy still exceeds 0.800. The advantage of ordinal random forest is larger measured by observations than by drugs. The superior performance of ordinal random forest over ordinal logistic regression implies the strong nonlinearity in the wedge dataset. We also note that the model performance measured by drugs is generally better than by observations on both datasets. This demonstrates the benefit of replication in the experimental design which reduces data noise in preclinical drug proarrhythmic assessment.[6,14,18] Interestingly, if we compare the same model across the two datasets, the model trained on the wedge dataset provides higher accuracy than on the stem cell dataset. We suspect that this is due to the experimental design of single site and platform in the wedge dataset, which likely introduces less data noise than the design of multiple sites and platforms in the stem cell dataset. However, the better model performance on the wedge dataset may overestimate the real-world prediction accuracy on other wedge datasets generated from various sites and platforms.

Supplementary Tables S5 – S8 show the confusion matrices of drug prediction by ordinal logistic regression and ordinal random forest. Among the wrong predictions, the ordinal logistic regression tends to underestimate the drug risk, while the ordinal random forest is more likely to overestimate the drug risk. On the stem cell dataset, the ordinal logistic regression underestimated six drugs' risks and overestimated four drugs' risks (Supplementary Table S5). In contrast, the ordinal random forest underestimated five drugs' risks and overestimated six drugs' risks (Supplementary Table S6). A similar pattern is observed on the wedge dataset (Supplementary Tables S7 and S8).

Overall, the ordinal logistic regression is a more aggressive model, and the ordinal random forest is a more conservative model in the prediction of drug TdP risk. We also find that all wrong predictions are between two adjacent risk categories, that is, high/intermediate or low/intermediate. There is no model predicting high risk as low risk or vice versa, thus avoiding the two more serious mistakes.

To examine the benefits of utilizing ordinal information in the drug TdP risk, we compared the three-category prediction accuracy of ordinal logistic regression, ordinal random forest, multinomial logistic regression, multinomial random forest, and classical ordinal logistic regression (ordered logit model) [62] (Supplementary Tables S9 and S10). The multinomial logistic regression and multinomial random forest treat the prediction of drug TdP risk as an ordinary classification task without using ordinal information. On the stem cell dataset, the ordinal random forest achieved the highest prediction accuracy measured by observations. If measured by drugs, the ordinal logistic regression and multinomial logistic regression exhibited the same highest prediction accuracy (Supplementary Table S9). On the wedge dataset, the ordinal random forest outperformed other models measured by both observations and drugs (Supplementary Table S10). The comparison result shows that the ordinal models outperform their counterparts without using ordinal information in most scenarios. We also find the ordinal logistic regression and ordinal random forest are more accurate than the classical ordinal logistic regression. We suspect there are two reasons for this observation. First, the classical ordinal logistic regression requires the proportional odds assumption, which the two ordinal models do not need. We conducted brand test [63] to examine this assumption and rejected the null hypothesis (the proportional odds assumption holds) at a 1% significance level on the stem cell dataset but failed to reject the null hypothesis at a 5% significance level on the wedge dataset. This explains the better performance of ordinal logistic regression versus classical ordinal logistic regression on the stem cell dataset and their similar performance on the wedge dataset. Second, the ordinal random forest is more capable of capturing the nonlinear relationship between predictors and TdP risk than any form of (linear) logistic regressions, including the classical ordinal logistic regression. This advantage explains the superiority of ordinal random forest on the wedge dataset, where nonlinear relation is likely crucial in the prediction of drug TdP risk. Additionally, the ordinal modeling used in this study provides more flexibility than the classical logistic regression. The two binary classifiers $\hat{f}_1$ and $\hat{f}_2$ could be substituted by appropriate classification algorithms to accommodate datasets generated from different *in vitro* studies.

*Model uncertainty measurement*

The four measurements in Table 1 and Supplementary Table S4 are point estimates of the model performance. To further understand the uncertainty of the model prediction, we conducted a stratified bootstrap[33,54] to construct the empirical distributions of the four model performance measurements. In each step of LODO-CV, we resampled with replacement separately on each drug in the training set to generate a stratified bootstrap sample. Then we trained a predictive model on this resampling training set and calculated its prediction performance on the test drug. For example, suppose that the current test drug is drug 28 in the stem cell dataset. Since there are 15 observations per drug, we resample 15 observations with replacement for each training drug from 1 to 27. Then we train the predictive model on this resampling dataset and measure its performance on the 15 observations of drug 28 (Figure 1). Our stratified bootstrap samples preserve the number of repeated observations for each drug as in the original dataset, and thus can be used to unbiasedly evaluate the uncertainty of model performance. We generated 1000 stratified bootstrap samples and repeated the previous model training and evaluation process to obtain the empirical distributions of three-category prediction accuracy, AUROC of high risk vs. intermediate-or-low risk, AUROC of high-or-intermediate risk vs. low risk, and concordance index.

Figure 3, Supplementary Figure S1, Table 2, and Supplementary Table S11 compare the model uncertainty of ordinal logistic regression and ordinal random forest. On the stem cell dataset, if measured by observations, the ordinal logistic regression has higher AUROC of high-or-intermediate risk vs. low risk. On the other hand, the ordinal random forest shows advantages in terms of three-category prediction accuracy and AUROC of high risk vs. intermediate-or-low risk. The concordance index is similar between the two models. If measured by drugs, the ordinal logistic regression has higher three-category prediction accuracy and concordance index, while the random forest shows advantages in terms of the two AUROCs. The two models also show comparable prediction variability. Overall, the comparison of model uncertainty on the stem cell dataset is consistent with the point estimate in Table 1 and Supplementary Table S4 — the ordinal logistic regression and ordinal random forest exhibit similar prediction performance.

On the wedge dataset, consistent with the point estimate in Table 1 and Supplementary Table S4, ordinal random forest consistently outperforms ordinal logistic regression in terms of all four measurements. The only exception is the similar three-category prediction accuracy measured by drugs between two models. The advantage of ordinal random forest is more obvious if measured by observations than by drugs. Like the point estimate, the same model trained on the wedge dataset provides higher accuracy than on the stem cell dataset. We also note that the variability of model prediction is lower for ordinal random forest than ordinal logistic regression. Again, the high and stable performance of random forest implies strong nonlinearity in the wedge dataset.

*Drug prediction analysis*

We calculated the proportion of correct predictions (correct rate) for each drug in the 1000 stratified bootstrap predictions. Figure 4 shows the correct rate of predicting each drug across different model-dataset combinations. In the high-risk group, bepridil and disopyramide have close-to-zero correct rates under both models on the stem cell dataset. However, they are better predicted in the wedge dataset, especially by the ordinal logistic regression. In the intermediate-risk group, chlorpromazine and clozapine have close-to-zero correct rates under both models in the stem cell dataset. On the other hand, cisapride and domperidone have low correct rates under both models in the wedge dataset. In the low-risk group, metoprolol and ranolazine have close-to-zero correct rates under both models in the stem cell dataset, while all drugs are predicted relatively well in the wedge dataset, except for ranolazine by the ordinal random forest.

Figure 4 also demonstrates the different asymptotic drug predictions on the two datasets. In the stem cell dataset, there are eight drugs on which both models resulted in less than 25% correct rates. These drugs with low prediction accuracy cover all three risk categories. In the wedge dataset, however, only two drugs of intermediate risk are difficult for both models to predict. If compared by models, ordinal logistic regression predicted nine drugs in the stem cell dataset and two drugs in the wedge dataset less than 25% correctly. On the other hand, ordinal random forest predicted 10 drugs in the stem cell dataset and five drugs in the wedge dataset less than 25% correctly. Although ordinal random forest shows better performance on the wedge dataset and similar performance on the stem cell dataset, it made more "absolute wrong predictions" asymptotically. The better performance of ordinal random forest measured by the previous four measurements is largely due to its capacity of predicting more drugs with high correct rates.

Many of the drugs with close-to-zero correct rates have been reported to have abnormal observations in their original experiments.[6,14] In the experiment of the stem cell dataset, for example, the high-risk drug bepridil surprisingly introduced no arrhythmias, a key indicator of high TdP risk, in hiPSC-CMs even at 30-fold of free clinical maximum concentration (Cmax). Another high-risk drug, disopyramide, was reported as being poorly soluble at the required concentrations, potentially biasing the prediction result. The two low-risk drugs with low correct rates, metoprolol and ranolazine, introduced abnormally strong signals of high TdP risk in hiPSC-CMs, including significant repolarization prolongation and arrhythmias. In the experiment of the wedge dataset, the intermediate-risk drug domperidone produced statistically significant QT prolongation at Cmax, one indicator of high TdP risk. Such observations imply that the wrong model predictions are possibly caused by an abnormality in the experiments rather than model incompetency.

*Sensitivity analysis*

We define the drugs with a correct rate of less than 25% under both models as potential outlier drugs.[6,53] Based on the drug prediction analysis shown in Figure 4, we identified eight and two potential outlier drugs in the stem cell dataset and wedge dataset, respectively. Table 3 shows those drugs and their average correct rates and wrong rates across the two models. On the stem cell dataset, the two models underestimated five outlier drugs' risks and overestimated three. On the wedge dataset, the two models overestimated two outlier drugs' risks. Similar to the point estimation of model performance (Supplementary Tables S5 – S8), the statistical models tend to make more aggressive prediction on the stem cell dataset and more conservative predictions on the wedge dataset.

To examine the effects of potential outlier drugs on the model performance, we removed them from the original datasets and redid the previous modeling and validation. Figure 5 and Supplementary Figure S2 compare the model performance before and after removing potential outlier drugs. First, we find that all four measurements — three-category prediction accuracy, AUROC of high risk vs. intermediate-or-low risk, AUROC of high-or-intermediate risk vs. low risk, and concordance index, improve without potential outlier drugs in the dataset, regardless of model, measured by observations or drugs, or dataset. Second, the benefit of removal is stronger in the stem cell dataset, partly because the baseline performance is already high in the wedge dataset. Third, the ordinal logistic regression and ordinal random forest have a similar performance boost after removing potential outlier drugs. Fourth, the absolute values of four measurements without potential outlier drugs are high, with many of them greater than 0.900, indicating excellent model predictive capacity. Overall, the existence of potential outlier drugs has significant impacts on the model performance, and most wrong predictions are caused by such drugs.

There are several common characteristics in the potential outlier drugs. First, many potential outlier drugs induced electrophysiological responses inconsistent with their true TdP risks. Those unusual responses were captured by statistical models and may have caused wrong predictions. Such drugs include bepridil, chlorpromazine, clozapine, domperidone, ranolazine, and metoprolol [6,14]. Second, some potential outlier drugs exhibited solubility issues in the experiments of stem cell dataset, including domperidone, azimilide, and disopyramide [6]. The solubility issues were likely to add noise to the electrophysiological responses, and thus lowered the data quality. The only potential outlier drug without inconsistent electrophysiological responses or significant solubility issue is cisapride in the wedge dataset. Third, all potential outlier drugs were falsely predicted to a TdP risk adjacent to the truth, that is, between low and intermediate or high and intermediate (Table 3). The models did not predict high risk as low risk or vice versa, the two more serious mistakes. Forth, two potential outlier drugs, bepridil and ranolazine, have also been identified in previous *in vitro* studies [6,55,56].

*Control analysis*

Motivated by the practice in the experimental assessment of drug-induced TdP risk,[34,35] we conducted control analysis to further validate the model prediction performance. Control drugs have clearly understood TdP risk; in the control analysis, the model trained on the training drugs will predict the test drug and control drug simultaneously. The predictive model is expected to correctly identify the risk of the control drug based on the data generated from the experiment. If the model prediction of the control drug is incorrect, then its simultaneous prediction on the test drug would be considered unreliable. We trust the model and its prediction of the test drug only if it can correctly predict the control drug. We select the high-risk drug sotalol as an example control drug in our analysis because of the thorough exploration of its mechanism in the literature[36] and its high correct rate in our drug prediction analysis (Figure 4). Any other drugs with definite TdP risk and high prediction accuracy can also be selected as the control drug.

Figure 6 shows the design of the control analysis. As with previous analyses, we trained ordinal logistic regression and ordinal random forest on the stem cell dataset and wedge dataset and validated the model performance using LODO-CV. In each step, the model was trained on a set with 26 drugs and predicted the risk of one left-out test drug and one pre-selected control drug (sotalol) simultaneously. The four model performance measurements were calculated for each test drug, conditioned on the correct prediction of control drug (with control) and no condition (without control). All measurements were calculated by observations and drugs. Table 4 and Supplementary Table S12 compare the model performance with and without control across all model-dataset combinations. Among the 16 measurements in the stem cell dataset, most of them with control are higher than (13) or equal to (2) their counterparts without control, except for the AUROC of high-or-intermediate risk vs. low risk under ordinal random forest by drugs. Among the eight measurements of ordinal logistic regression in the wedge dataset, most of them are higher than (6) or equal to (1) their counterparts without control, except for the AUROC of high-or-intermediate risk vs. low risk by observations. Ordinal random forest has the same performance regardless of control because the prediction accuracy of the control drug is 100%. The result of control analysis illustrates that the prediction of a control drug is an effective indicator of model capacity — the models that can correctly predict control drugs generally perform better than those that cannot. In practice, we may consider the model outputs conditioned on the correct prediction of the control drug as high-confidence results and other outputs as low-confidence results. It should be noted that there must be appropriate control drugs in the original dataset to conduct the control analysis.

*Predictor importance analysis*

To examine the effect of various predictors on the model performance, we conducted a predictor importance analysis. We utilized the permutation predictor importance[25,37] to measure the contribution of each predictor to the model prediction. The permutation predictor importance of one predictor is defined as the decrease of the model prediction accuracy when that predictor's value is randomly shuffled. Since the random shuffling breaks the relationship between the predictor and target variable (TdP risk), the decrease of the prediction accuracy (if any) indicates the model dependency on that shuffled predictor. Using permutation predictor importance has three advantages. First, the calculation of permutation predictor importance is independent of the specific model form. Second, the predictive model only needs to be trained once. Third, the random shuffling can be repeated multiple times to reduce the variability in the calculation.

Suppose there are $P$ predictors in dataset $D$, and $\hat{f}$ is a pretrained predictive model. Let $acc$ be the upper bound of 95% confidence interval of three-category prediction accuracy measured by observations (Supplementary Table S11). For each predictor $j$ ($j$th column of $D$) and each repetition $g$ in $1, \ldots, G$, we perform the following calculations:

1. Randomly shuffle the $j$th column of dataset $D$ to generate a permuted dataset $\widetilde{D_{g,J}}$

2. Calculate the three-category prediction accuracy $acc_{g,j}$ on the permuted dataset $\widetilde{D_{g,J}}$

Then the permutation importance of predictor $j$, $imp_j$, is calculated as

$$imp_j = acc - \frac{1}{G}\sum_{g=1}^{G} acc_{g,j}$$

Finally, we calculate the normalized permutation predictor importance of predictor $j$, $nimp_j$, by

$$nimp_j = \frac{imp_j}{max\ \{imp_1, imp_2, \ldots, imp_J\}}$$

In our implementation, we set the repetition number $G$ to 100 and $\hat{f}$ to ordinal random forest.

Figure 7 shows the normalized permutation predictor importance for all seven predictors in the stem cell dataset and top-seven predictors in the wedge dataset. In the stem cell dataset, the top-two important predictors in terms of contribution to prediction accuracy are 1) the drug concentration when drug-induced arrhythmias were first observed (foldaym); 2) whether drug-induced arrhythmias occurred at any concentration in more than 40% wells (AMtwooutoffive). These two predictors also lead to a large margin of importance over other predictors. This result is consistent with the previous study showing that the introduction of arrhythmias is the most important factor in the identification of drug-induced TdP risk on the hiPSC-CMs protocol.[9,38]

In the wedge dataset, the top-two important predictors are 1) interval between the J point and the start of T wave (c3v2); 2) interval between the J point and the peak of T wave (c3v3). They also echo the previous findings in the experiments of RVWA.[11,12] The consistency between model discovery and established drug-risk mechanism further validates the model specification and provides another insight into the model interpretability. It should be mentioned that the permutation predictor importance emphasizes the marginal effects of predictors on the model prediction.[39] Therefore, the non-top important predictors in Figure 7 may also impact the model performance through their interactions with other predictors.

We note that the QT interval (c3v9) is not among the top-seven most important predictors, but the JT interval (c3v2) is. There are two potential reasons for this result. First, several studies have found that JT interval and QT interval exhibit similar predictive power in the assessment of ventricular repolarization, coronary heart disease (CHD), and mortality caused by TdP.[57,58] Under certain circumstances, JT interval can show prognostic advantage compared with QT interval, especially if the QS interval is prolonged.[59] Second, the permutation predictor importance is model specific. It is calculated based on the better-performed model, ordinal random forest, on the wedge dataset. Other types of models with similar performance may generate different predictor importance. Additionally, the permutation predictor importance may not capture the between-predictor interactions. Therefore, the model-based predictor importance should be treated as one reference rather than the conclusion of true predictor importance.

**Discussion**

The overall model performance, uncertainty measurement, and drug prediction demonstrate that the prediction accuracy of two models on the wedge dataset is consistently higher than the stem cell dataset. Such discrepancy is largely due to the different experimental designs of the two datasets. Observations in the stem cell dataset were generated at 10 experimental sites, using two hiPSC-CM lines and five EP platforms. On the other hand, all observations in the wedge dataset were generated at one laboratory using the same type of biological sample and EP platform. Although the multisite experiments were supposed to follow the same protocols, the batch effects caused by site-to-site variability introduced a higher degree of noise in the stem cell dataset. The signal-noise ratio in the stem cell dataset is thus lower than the wedge dataset, resulting in lower prediction accuracy. One potential solution for this issue is to estimate the effects of site, cell line, and EP platform and include these factors into the modeling process. The accurate estimation of such effects requires special experimental designs with enough power to describe and explain the variations from those factors.[40–43]

The objective of CiPA and RVWA paradigms is to improve the drug TdP risk assessment by measuring variables beyond only hERG current block or QT prolongation. To evaluate the advantage of this strategy, we compare the three-category prediction accuracy between multivariate models and univariate models (Supplementary Figure S3). The multivariate models utilize all predictors in the dataset, while the univariate models only contain predictor arrhythmia (on the stem cell dataset) or QT interval (c3v9, on the wedge dataset), two commonly used variables in the univariate study. The comparison shows a significant accuracy improvement of multivariate models over their univariate counterparts. The multivariate ordinal logistic regression even doubles the accuracy on the stem cell dataset over its univariate version measured by both observations and drugs. The comparison result further validates the benefits of combining *in vitro* multivariate studies and *in silico* statistical learning models.

Besides the CiPA and RVWA paradigms, there are many *in vitro* protocols aiming to assess the drug-induced TdP risk, including automated patch clamp,[44] human ventricular myocyte model,[15] rabbit isolated hearts,[45] *in vitro* atrioventricular block dog,[16] and others. It is difficult to fairly compare those protocols because they were developed under various biological models, experimental conditions, methods of data collection, and predictor formulation. One future work is to utilize our statistical learning framework to examine the strength of these protocols in the assessment of drug-induced TdP risk. Here, the strength of protocols can be defined as the quality of datasets generated from experiments. A well-designed *in vitro* protocol should generate data including sufficient information and minimum noise for *in silico* models to accurately predict drug TdP risk. A statistical learning model, e.g., an ordinal random forest, can be trained on datasets generated by different protocols. The model prediction accuracy across different datasets will serve as a proxy of the protocol's strength. In addition, the model interpretation can be used as another metric to compare the strength of protocols. A good *in vitro* study should generate datasets on which the model interpretation is consistent with domain knowledge.

The second future work is to utilize the proposed statistical learning framework to find new predictors from the entire ECG to improve *in vitro* protocols. Although the selection of predictors is well established in the previous studies of CiPA initiative and rabbit ventricular wedge,[10,60] there might be informative variables that were ignored from the original *in vitro* studies. Potential predictors suggested by domain experimentalists can be used to fit statistical learning models and new predictors with better model performance would be included in the existing *in vitro* protocols.

Another future direction is to encompass novel experimental techniques and measurements beyond electrophysiological signals into the *in vitro* paradigm. For example, single-cell RNA-sequencing (scRNA-seq),[46] a next-generation sequencing technology revealing the genome-wide gene expression at single-cell levels, can be applied to evaluate and compare the transcriptomes of individual cells before and after drug treatment.[47–49] The gene expression dynamic associated with drug-induced TdP risk will potentially improve the assessment of such risks, similar to the progress made by scRNA-seq in cancer diagnosis, cell type discovery, and precision medicine.[50,51]

In summary, we proposed a comprehensive statistical modeling framework to predict drug-induced TdP risk from two preclinical experimental protocols. Ordinal logistic regression and ordinal random forest were trained on the stem cell dataset and wedge dataset. We calculated the unbiased estimate of model performance by LODO-CV. The uncertainty of model performance was evaluated by stratified bootstrap. Our proposed method provided interpretability consistent with domain knowledge through normalized permutation predictor importance. Sensitivity analysis identified potential outlier drugs that were also described in the literature. Finally, we conducted a control analysis and further validated the model performance. Our work is a valuable addition to the current attempts to construct trustful *in silico* models in the assessment of drug-induced TdP risk.

**Acknowledgments**

We thank Dr. Gan-Xin Yan for sharing the wedge dataset with us. We also thank the Health and Environmental Sciences Institute (HESI) for providing the stem cell dataset from the hiPSC-CMs study funded by FDA Broad Agency Announcement (BAA) contract (FDABAA-15-00121). We are grateful for the valuable discussions with Drs. Christine Garnett, Donglin Guo, Lars Johannesen, Jose Vicente Ruiz, and Wendy Wu. Finally, we want to thank Drs. Yi Tsong and Atiar Rahman for their support for this project.

**References**

1. Al-Khatib, S. M., Stevenson, W.G., Ackerman, M. J., et al. 2018. 2017 AHA/ACC/HRS guideline for management of patients with ventricular arrhythmias and the prevention of sudden cardiac death: Executive summary: A report of the American college of cardiology/American heart association task force on clinical practice guidelines and the heart rhythm society. *Journal of the American College of Cardiology* **72**, 1677–1749.
2. Joshi, A., Dimino, T., Vohra, Y., Cui, C. & Yan, G.-X. 2004. Preclinical strategies to assess QT liability and torsadogenic potential of new drugs: the role of experimental models. *Journal of Electrocardiology* **37** Suppl, 7–14.
3. Stockbridge, N., Morganroth, J., Shah, R. R. & Garnett, C. 2013. Dealing with global safety issues : was the response to QT-liability of non-cardiac drugs well coordinated? *Drug Safety* **36**, 167–182.
4. Cavero, I. & Crumb, W. 2005. ICH S7B draft guideline on the non-clinical strategy for testing delayed cardiac repolarisation risk of drugs: a critical analysis. *Expert Opinion on Drug Safety* **4**, 509–530.
5. Shah, R. R. 2005. Drugs, QTc interval prolongation and final ICH E14 guideline: an important milestone with challenges ahead. *Drug Safety* **28** 1009–1028.
6. Blinova, K. et al. International multisite study of human-induced pluripotent stem cell-derived cardiomyocytes for drug proarrhythmic potential assessment. *Cell Reports* **24**, 3582–3592 (2018).
7. Badri, M. et al. Mexiletine prevents recurrent torsades de pointes in acquired long QT syndrome refractory to conventional measures. *JACC Clinical Electrophysiology* **1**, 315–322 (2015).
8. Fermini, B. et al. 2016. A new perspective in the field of cardiac safety testing through the comprehensive in vitro proarrhythmia assay paradigm. *Journal of Biomolecular Screening* **21**, 1–11.
9. Vicente, J. et al. 2018. Mechanistic model-informed proarrhythmic risk assessment of drugs: Review of the "CiPA" initiative and design of a prospective clinical validation study. *Clinical Pharmacology & Therapeutics* **103**, 54–66.
10. Yan, G. X. et al. 2001. Ventricular hypertrophy amplifies transmural repolarization dispersion and induces early afterdepolarization. *American Journal of Physiology Heart and Circulatory Physiology* **281**, H1968-75.
11. Wang, D., Patel, C., Cui, C. & Yan, G.-X. 2008. Preclinical assessment of drug-induced proarrhythmias: role of the arterially perfused rabbit left ventricular wedge preparation. *Pharmacology & Therapeutics* **119**, 141–151.
12. Liu, T., Brown, B.S., Wu, Y., Antzelevitch, C., Kowey, P.R. & Yan, G.-X. 2006. Blinded validation of the isolated arterially perfused rabbit ventricular wedge in preclinical assessment of drug-induced proarrhythmias. *Heart Rhythm* **3**, 948–956.
13. Liu, T. *et al.* 2013. Differentiating electrophysiological effects and cardiac safety of drugs based on in vitro electrocardiogram: A blinded validation. *Journal of Pharmacological and Toxicological Methods* **68**, e23.
14. Liu, T. *et al.* 2021. Utility of Normalized TdP Score System in drug proarrhythmic potential assessment: A blinded in vitro study of CiPA drugs. *Clinical Pharmacology & Therapeutics* 109, 1606–1617.
15. Lancaster, M. C. & Sobie, E. A. 2016. Improved Prediction of Drug-Induced Torsades de Pointes Through Simulations of Dynamics and Machine Learning Algorithms. *Clinical Pharmacology & Therapeutics* **100**, 371–379.
16. Passini, E. et al. 2017. Human in silico drug trials demonstrate higher accuracy than animal models in predicting clinical pro-arrhythmic cardiotoxicity. *Frontiers in Physiology* 8: 668.
17. Okada, J.-I. et al. 2015. Screening system for drug-induced arrhythmogenic risk combining a patch clamp and heart simulator. *Science Advances* **1**, e1400142.

18. Li, Z. et al. 2019. Assessment of an in silico mechanistic model for proarrhythmia risk prediction under the CiPA initiative. *Clinical Pharmacology & Therapeutics* **105**, 466–475.
19. Hastie, T., Tibshirani, R. & Friedman, J. 2009. *The Elements of Statistical Learning: Data Mining, Inference, and Prediction, Second Edition*. New York: Springer Science & Business Media.
20. Breiman, L. 1996. Bagging predictors. *Machine Learning* **24**, 123–140.
21. Kuhn, M. 2008. Building Predictive Models in R Using the caret Package. *Journal of Statistical Software, Articles* **28**, 1–26.
22. Cardoso, J. S. & Sousa, R. 2011. Measuring the performance of ordinal classification. *International Journal of Pattern Recognition and Artificial Intelligence* **25**, 1173–1195.
23. Gaudette, L. & Japkowicz, N. 2009. Evaluation Methods for Ordinal Classification. in *Advances in Artificial Intelligence* 207–210.
24. Frank, E. & Hall, M. 2001. A Simple Approach to Ordinal Classification. in *Machine Learning: ECML 2001* 145–156.
25. Breiman, L. Random Forests. *Machine Learning* **45**, 5–32 (2001).
26. Liaw, A., Wiener, M. Classification and regression by randomForest. *R news* **2**, 18–22 (2002).
27. Couronné, R., Probst, P. & Boulesteix, A.-L. Random forest versus logistic regression: a large-scale benchmark experiment. *BMC Bioinformatics* **19**, (2018).
28. Grau, J., Grosse, I. & Keilwagen, J. 2015. PRROC: computing and visualizing precision-recall and receiver operating characteristic curves in R. *Bioinformatics* **31**, 2595–2597.
29. Raykar, V. C., Steck, H., Krishnapuram, B., Dehing-Oberije, C. & Lambin, P. 2007. On ranking in survival analysis: bounds on the concordance index. in *Proceedings of the 20th International Conference on Neural Information Processing Systems* 1209–1216 (Curran Associates Inc., 2007).
30. Allaire, J. J. *et al.* reticulate: Interface to'Python'. *R package version* **1**, (2018).
31. Davidson-Pilon, C. 2019. lifelines: survival analysis in Python. *Journal of Open Source Software* **4**, 1317.
32. R Core Team 2013. R: A language and environment for statistical computing. *R Foundation for statistical computing, Vienna*.
33. Mashreghi, Z., Haziza, D. & Léger, C. 2016. A survey of bootstrap methods in finite population sampling. *Statistics Surveys,* **10**, 1–52.
34. Gintant, G. 2011. An evaluation of hERG current assay performance: Translating preclinical safety studies to clinical QT prolongation. *Pharmacology & Therapeutics* **129**, 109–119.
35. Rampe, D. & Brown, A. M. 2013. A history of the role of the hERG channel in cardiac risk assessment. *Journal of Pharmacological and Toxicological Methods* **68**, 13–22.
36. Singh, B. N., Kehoe, R., Woosley, R. L., Scheinman, M. & Quart, B. 1995. Multicenter trial of sotalol compared with procainamide in the suppression of inducible ventricular tachycardia: A double-blind, randomized parallel evaluation. *American Heart Journal* **129**, 87–97.
37. Altmann, A., Toloşi, L., Sander, O. & Lengauer, T. 2010. Permutation importance: a corrected feature importance measure. *Bioinformatics* **26**, 1340–1347.
38. Colatsky, T. *et al.* 2016. The Comprehensive in Vitro Proarrhythmia Assay (CiPA) initiative - Update on progress. *Journal of Pharmacological and Toxicological Methods* **81**, 15–20.
39. Strobl, C., Boulesteix, A.-L., Kneib, T., Augustin, T. & Zeileis, A. 2008. Conditional variable importance for random forests. *BMC Bioinformatics* **9**, 307.
40. Wang, L., Xiao, Q. & Xu, H. 2018. Optimal maximin $L_1$-distance Latin hypercube designs based on good lattice point designs. *Annals of Statistics* 46 (6B), 3741-3766.

41. Xiao, Q., Wang, L. & Xu, H. 2019. Application of kriging models for a drug combination experiment on lung cancer. *Statistics in Medicine.* **38**, 236–246.
42. Wang, L., Yang, J.-F., Lin, D. K. J. & Liu, M.-Q. Nearly Orthogonal Latin Hypercube Designs For Many Design Columns. *Statistica Sinica* **25**, 1599–1612 (2015).
43. Wang, L., Sun, F., Lin, D. K. J. & Liu, M.-Q. 2018. CONSTRUCTION OF ORTHOGONAL SYMMETRIC LATIN HYPERCUBE DESIGNS. *Statistica Sinica* **28**, 1503–1520.
44. Kramer, J. *et al.* 2013. MICE Models: Superior to the HERG Model in Predicting Torsade de Pointes. *Scientific Reprts* **3**, 1–7 (2013).
45. Vargas, H. M. *et al*. 2015. Evaluation of drug-induced QT interval prolongation in animal and human studies: a literature review of concordance. *British Journal of Pharmacologyi* **172**, 4002–4011 (2015).
46. Hwang, B., Lee, J. H. & Bang, D. 2018. Single-cell RNA sequencing technologies and bioinformatics pipelines. *Experimental & Molecular Medicine* **50**, 96.
47. Haque, A., Engel, J., Teichmann, S. A. & Lönnberg, T. 2017. A practical guide to single-cell RNA-sequencing for biomedical research and clinical applications. *Genome Medicine.* **9**.
48. Xi, N. M. & Li, J. J. 2021. Benchmarking Computational Doublet-Detection Methods for Single-Cell RNA Sequencing Data. *Cell Systems* **12**, 176-194.e6.
49. Xi, N. M. & Li, J. J. 2021. Protocol for executing and benchmarking eight computational doublet-detection methods in single-cell RNA sequencing data analysis. *STAR Protocols 2(3):100699.*
50. Sun, G. *et al.* 2021. Single-cell RNA sequencing in cancer: Applications, advances, and emerging challenges. *Molecular Therapy - Oncolytics* **21**, 183–206.
51. Wiedmeier, J. E., Noel, P., Lin, W., Von Hoff, D. D. & Han, H. 2019. Single-Cell Sequencing in Precision Medicine. *Cancer Treatment and Research Communications* **178**, 237–252.
52. Christophe, B. 2015. In silico study of transmural dispersion of repolarization in non-failing human ventricular myocytes: Contribution to cardiac safety pharmacology. *British Journal of Pharmacology.* **7**, 88–101.
53. Yim, D.-S. 2018. Five years of the CiPA project (2013-2018): what did we learn? *Translational and clinical pharmacology*, 26, 145–149.
54. Shankar, P.M. 2020. Tutorial overview of simple, stratified, and parametric bootstrapping. *Engineering Reports*, 2(1), p.e12096.
55. Crumb, W.J., Jr., Vicente, J., Johannesen, L., and Strauss, D.G. 2016. An evaluation of 30 clinical drugs against the comprehensive in vitro proarrhythmia assay (CiPA) proposed ion channel panel. *Journal of Pharmacological and Toxicological Methods*, 81, 251–262.
56. Johannesen, L., Vicente, J., Mason, J.W., Erato, C., Sanabria, C., Waite-Labott, K., Hong, M., Lin, J., Guo, P., Mutlib, A., et al. 2016. Late sodium current block for drug-induced long QT syndrome: Results from a prospective clinical trial. *Clinical Pharmacology & Therapeutics* 99, 214–223.
57. Crow, R.S., Hannan, P.J. and Folsom, A.R., 2003. Prognostic significance of corrected QT and corrected JT interval for incident coronary heart disease in a general population sample stratified by presence or absence of wide QRS complex: the ARIC Study with 13 years of follow-up. *Circulation*, 108(16), pp.1985-1989.
58. Rautaharju, P.M., Zhang, Z.M., Prineas, R. and Heiss, G., 2004. Assessment of prolonged QT and JT intervals in ventricular conduction defects. *The American journal of cardiology*, 93(8), pp.1017-1021.
59. Zulqarnain, M.A., Qureshi, W.T., O'Neal, W.T., Shah, A.J. and Soliman, E.Z., 2015. Risk of mortality associated with QT and JT intervals at different levels of QRS duration (from the third national health and nutrition examination survey). *The American journal of cardiology*, 116(1), pp.74-78.

60. Ando, H., Yoshinaga, T., Yamamoto, W., Asakura, K., Uda, T., Taniguchi, T., Ojima, A., Shinkyo, R., Kikuchi, K., Osada, T. and Hayashi, S., 2017. A new paradigm for drug-induced torsadogenic risk assessment using human iPS cell-derived cardiomyocytes. *Journal of pharmacological and toxicological methods*, 84, pp.111-127.
61. Menard, S., 2002. Applied logistic regression analysis (Vol. 106). Sage.
62. McCullagh, P., 1980. Regression models for ordinal data. *Journal of the Royal Statistical Society: Series B (Methodological)*, 42(2), pp.109-127.
63. Brant, R., 1990. Assessing proportionality in the proportional odds model for ordinal logistic regression. *Biometrics*, pp.1171-1178.
64. Gutiérrez, P.A., Perez-Ortiz, M., Sanchez-Monedero, J., Fernandez-Navarro, F. and Hervas-Martinez, C., 2015. Ordinal regression methods: survey and experimental study. *IEEE Transactions on Knowledge and Data Engineering*, 28(1), pp.127-146.

**Figures**

|  | Train (27 drugs) | | | | | | | | Test |
|---|---|---|---|---|---|---|---|---|---|
| Step 1 | 1 | 2 | 3 | 4 | …… | 25 | 26 | 27 | 28 |
| Step 2 | 1 | 2 | 3 | 4 | …… | 25 | 26 | 28 | 27 |
| Step 3 | 1 | 2 | 3 | 4 | …… | 25 | 27 | 28 | 26 |
| ⋮ | | | | | ⋮ | | | | |
| Step 27 | 1 | 3 | 4 | 5 | …… | 26 | 27 | 28 | 2 |
| Step 28 | 2 | 3 | 4 | 5 | …… | 26 | 27 | 28 | 1 |

Measurement of model performance on 28 drugs

**Figure 1. The schematic diagram of leave-one-drug-out cross-validation.** In each iteration, we trained the predictive model on 27 training drugs and predicted one left-out drug. The same process was repeated until each drug was predicted, and the model performance was calculated by combining the prediction results of each left-out drug.

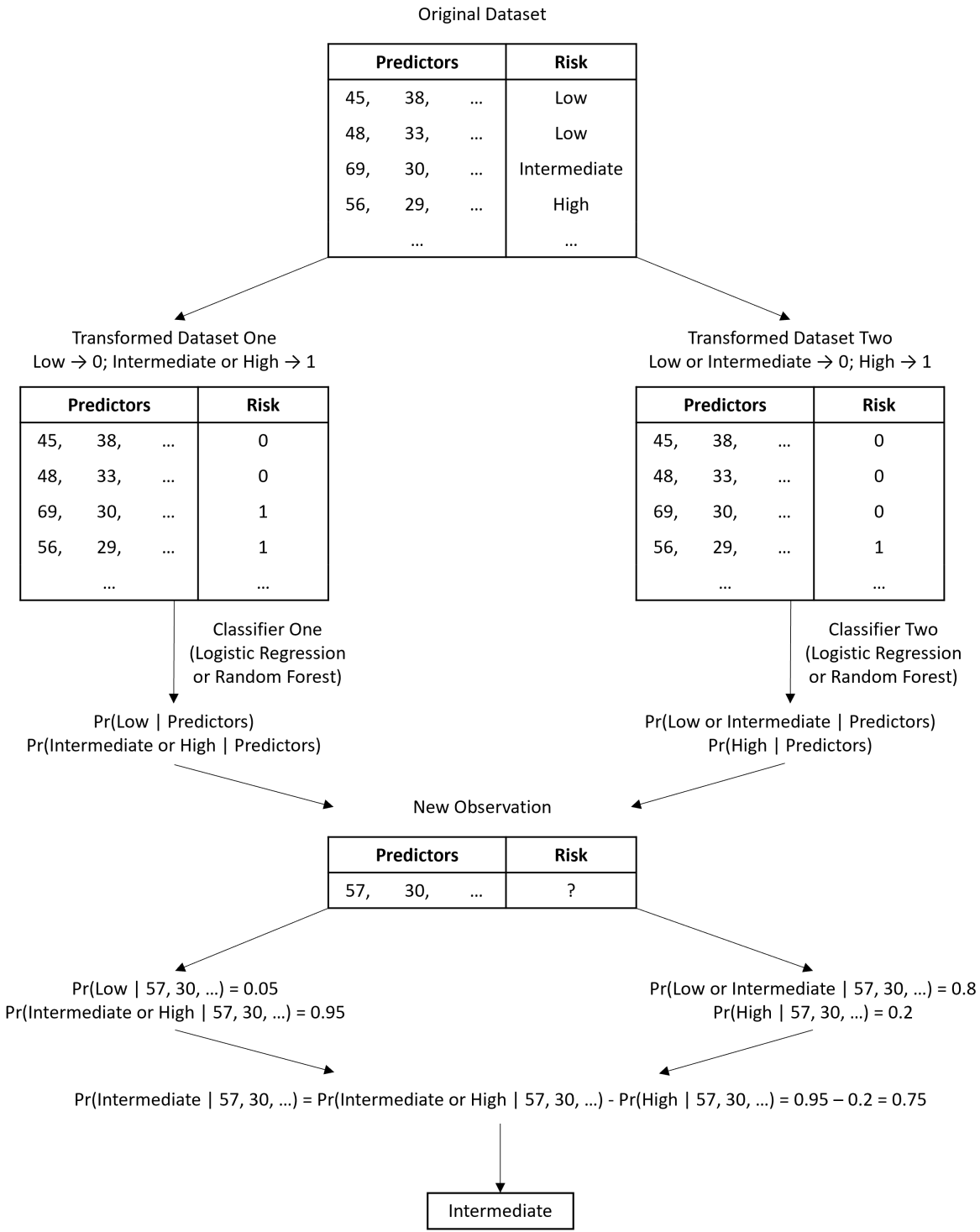

**Figure 2. The schematic diagram of utilizing ordinal information in the prediction of drug TdP risk.** First, the original dataset is transformed into two datasets, one with low risk labeled as 0 and intermediate and high risk labeled as 1, another with low and intermediate risk labeled as 0 and high risk labeled as 1. Second, two binary classifiers with the same type are trained on the two transformed datasets. Third, the two classifiers will output the probabilities of low risk, intermediate or high risk, low or intermediate risk, and high risk for any new observations. Fourth, the probability of intermediate risk is calculate as Pr(Intermediate or High) – Pr(High). Finally, the new observation will be predicted to the risk with the highest probability.

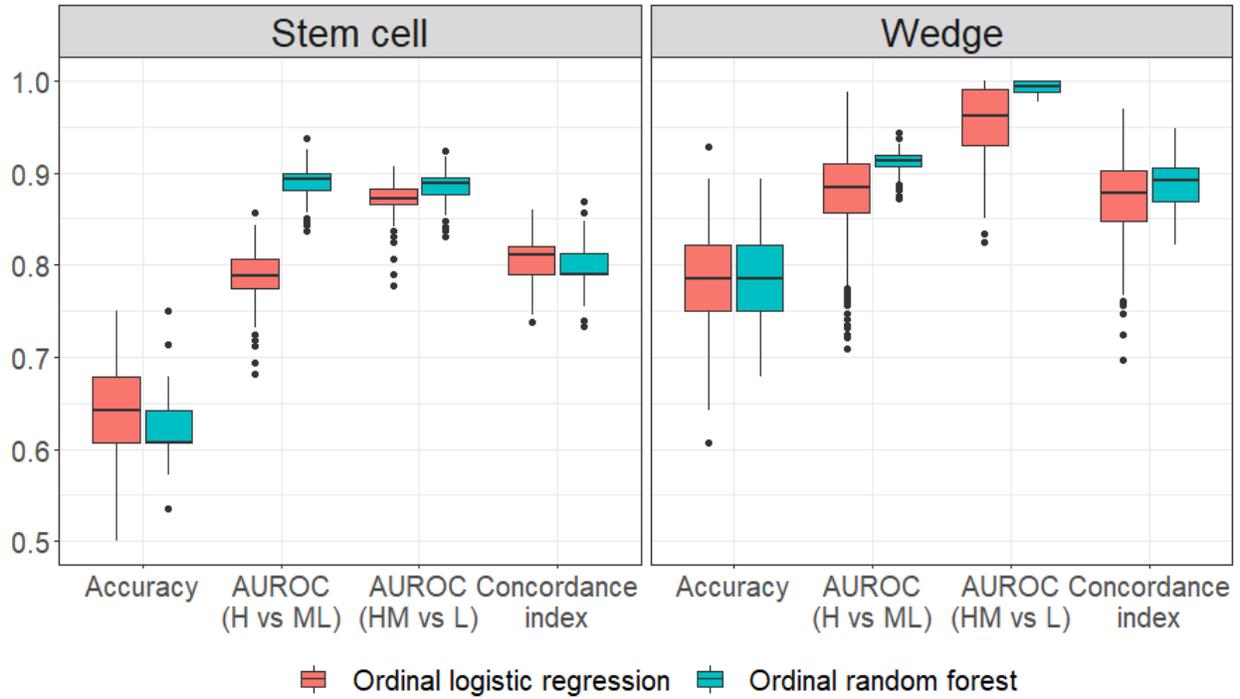

**Figure 3. The empirical distributions of model performance under stratified bootstrap measured by drugs.** Four performance measurements of two models were calculated on the stem cell and wedge datasets.

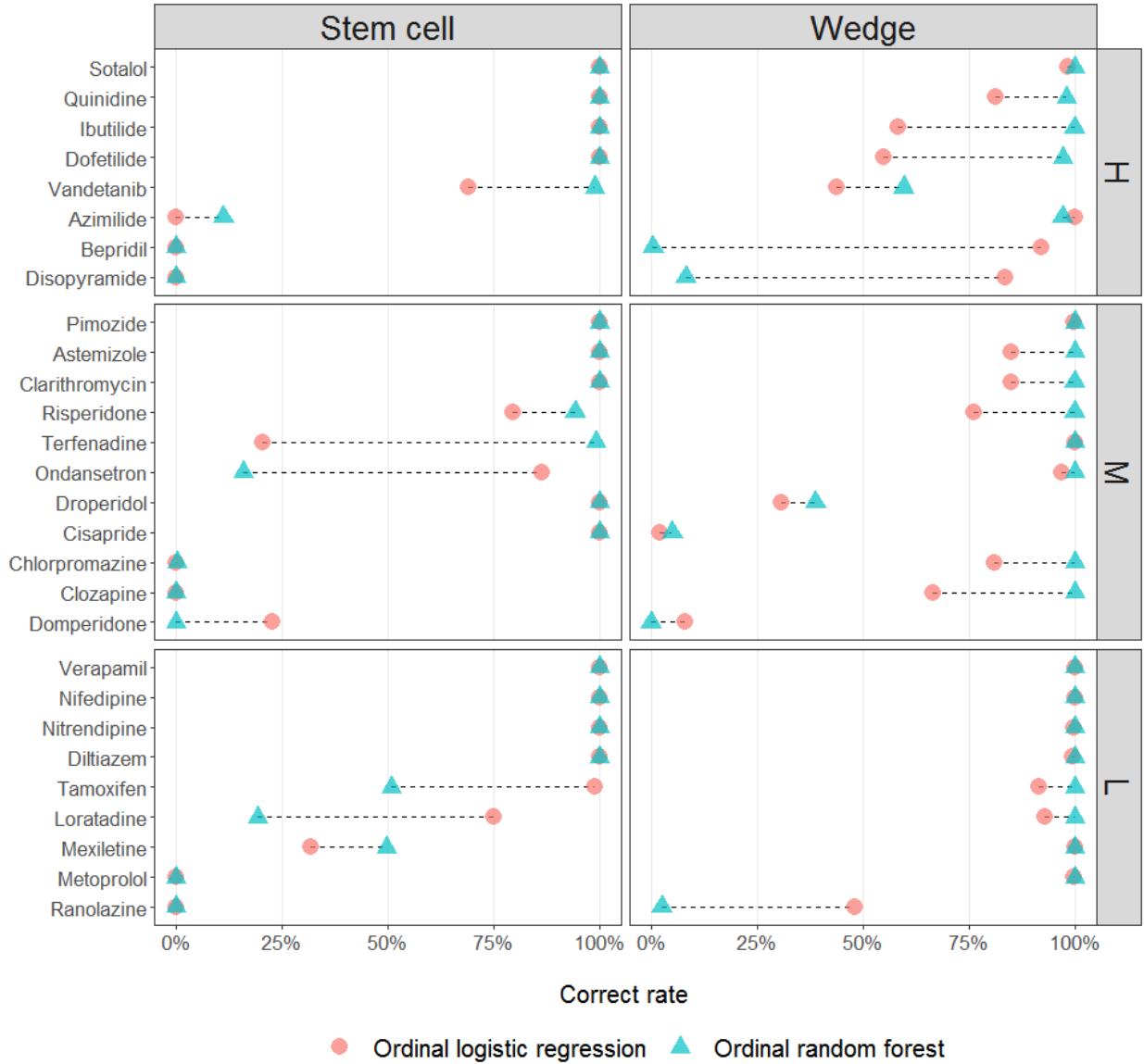

**Figure 4. The correct rate of drug prediction calculated by stratified bootstrap.** For each drug, we connect the correct rates of two models. In each risk category, drugs are sorted from high to low based on their average correct rates across two models.

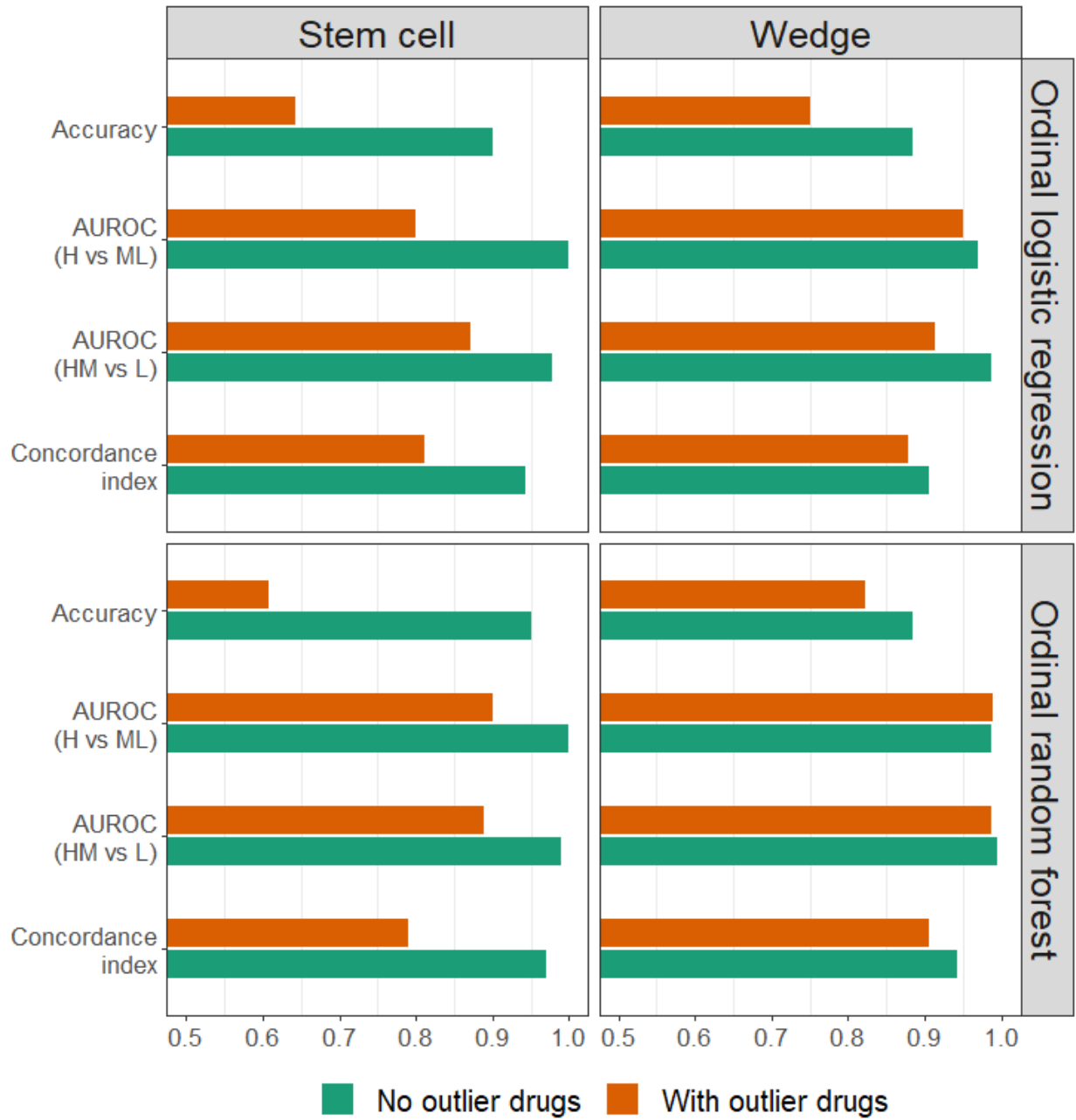

**Figure 5. The model performance before and after the removal of potential outlier drugs measured by drugs.** The four measurements were calculated under four conditions: 1) ordinal logistic regression on the stem cell dataset; 2) ordinal logistic regression on the wedge dataset; 3) ordinal random forest on the stem cell dataset; 4) ordinal random forest on the wedge dataset.

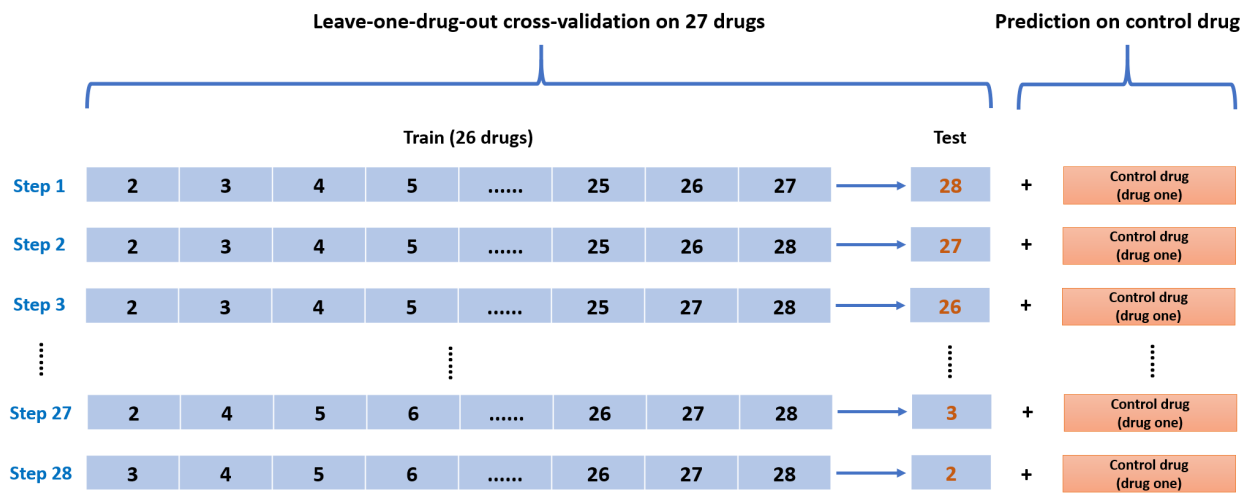

**Figure 6. The schematic diagram of control analysis where drug one is the example control drug.** We trained the model on 26 drugs and predicted one test drug and one control drug simultaneously. Leave-one-drug-out cross-validation was used to estimate the prediction accuracy of 27 non-control drugs and one control drug.

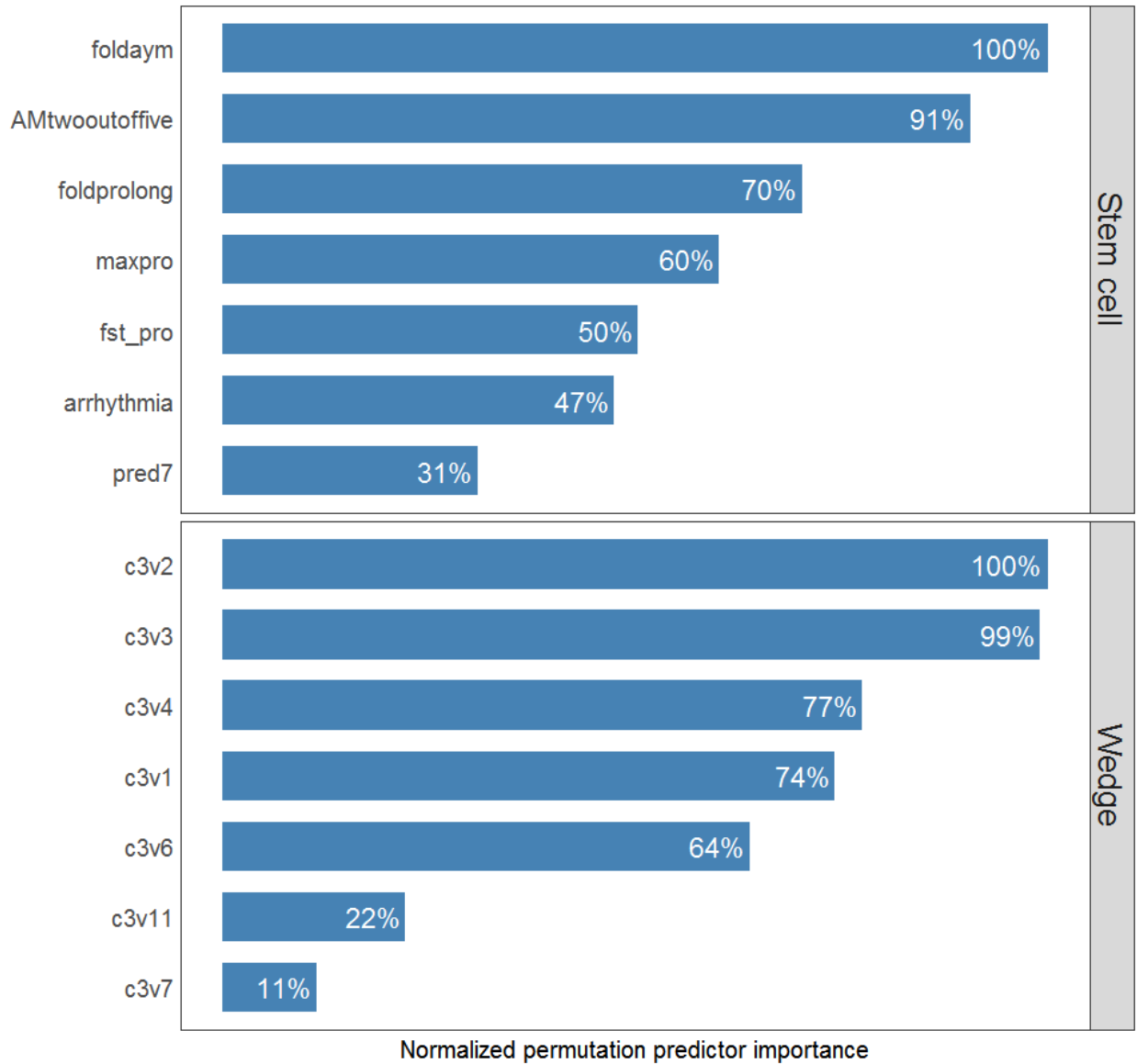

**Figure 7. The normalized permutation predictor importance on the stem cell dataset and wedge dataset**. Predictors are sorted from high to low. There are seven predictors in the stem cell dataset and the top-seven predictors in the wedge dataset.

# Tables

| Dataset | Model | Accuracy (Three-category) | AUROC (High vs. Intermediate-or-Low) | AUROC (High-or-Intermediate vs. Low) | Concordance index (Three-category) |
|---|---|---|---|---|---|
| Stem cell | Ordinal logistic regression | <u>0.643</u> | 0.800 | 0.872 | <u>0.811</u> |
| | Ordinal random forest | 0.607 | <u>0.900</u> | <u>0.889</u> | 0.790 |
| Wedge | Ordinal logistic regression | 0.750 | 0.950 | 0.913 | 0.879 |
| | Ordinal random forest | <u>0.822</u> | <u>0.989</u> | <u>0.988</u> | <u>0.906</u> |

**Table 1. The model performance measured by drugs (point estimate).** Three-category prediction accuracy, two AUROCs, and concordance index are shown for ordinal logistic regression and ordinal random forest. The higher values between the two models are underscored.

| Dataset | Model | Measurement | Accuracy (Three-category) | AUROC (High vs. Intermediate-or-Low) | AUROC (High-or-Intermediate vs. Low) | Concordance index (Three-category) |
|---|---|---|---|---|---|---|
| Stem cell | Ordinal logistic regression | 95% CI | (0.572, 0.715) | (0.750, 0.825) | (0.842, 0.901) | (0.763, 0.848) |
| | | Mean | <u>0.637</u> | <u>0.789</u> | 0.873 | <u>0.807</u> |
| | Ordinal random forest | 95% CI | (0.571, 0.679) | (0.857, 0.919) | (0.854, 0.913) | (0.767, 0.836) |
| | | Mean | 0.622 | 0.892 | <u>0.885</u> | 0.797 |
| Wedge | Ordinal logistic regression | 95% CI | (0.679, 0.893) | (0.779, 0.957) | (0.883, 1.000) | (0.784, 0.948) |
| | | Mean | 0.777 | 0.880 | 0.959 | 0.875 |
| | Ordinal random forest | 95% CI | (0.750, 0.858) | (0.888, 0.932) | (0.983, 1.000) | (0.860, 0.927) |
| | | Mean | <u>0.789</u> | <u>0.913</u> | <u>0.994</u> | <u>0.890</u> |

**Table 2. The summary statistics of model performance under stratified bootstrap measured by drugs.** The 95% confidence intervals and means of four measurements were calculated for ordinal logistic regression and ordinal random forest. The higher mean values between the two models are underscored.

| Dataset | Drug | True TdP risk (correct rate) | Predicted TdP risk (wrong rate) |
|---|---|---|---|
| Stem cell | Azimilide | High (0.056) | Intermediate (0.944) |
| | Bepridil | High (0.000) | Intermediate (1.000) |
| | Disopyramide | High (0.000) | Intermediate (0.999) |
| | Domperidone | Intermediate (0.114) | High (0.886) |
| | Chlorpromazine | Intermediate (0.002) | Low (0.995) |
| | Clozapine | Intermediate (0.000) | Low (1.000) |
| | Ranolazine | Low (0.000) | Intermediate (0.990) |
| | Metoprolol | Low (0.000) | Intermediate (1.000) |
| Wedge | Domperidone | Intermediate (0.040) | High (0.928) |
| | Cisapride | Intermediate (0.035) | High (0.965) |

**Table 3. The potential outlier drugs in the stem cell dataset and wedge dataset.** The potential outlier drugs are defined as drugs with both correct rates under two models less than 25% in stratified bootstrap. The correct rate is the proportion of correctly predicting the true TdP risk. The wrong rate is the proportion of most wrong predictions of TdP risk. They are both calculated by averaging the prediction results across two models.

| Dataset | Model | Measurement | Without control | With control |
|---|---|---|---|---|
| Stem cell | Ordinal logistic regression | Accuracy | <u>0.629</u> | <u>0.629</u> |
| | | Concordance index | 0.782 | <u>0.785</u> |
| | | AUROC (H vs. ML) | 0.750 | <u>0.765</u> |
| | | AUROC (HM vs. L) | 0.833 | <u>0.858</u> |
| | Ordinal random forest | Accuracy | 0.592 | <u>0.630</u> |
| | | Concordance index | 0.755 | <u>0.783</u> |
| | | AUROC (H vs. ML) | <u>0.893</u> | <u>0.893</u> |
| | | AUROC (HM vs. L) | <u>0.882</u> | 0.877 |
| Wedge | Ordinal logistic regression | Accuracy | 0.777 | <u>0.815</u> |
| | | Concordance index | 0.887 | <u>0.910</u> |
| | | AUROC (H vs. ML) | 0.907 | <u>0.929</u> |
| | | AUROC (HM vs. L) | <u>0.963</u> | <u>0.963</u> |
| | Ordinal random forest | Accuracy | <u>0.815</u> | <u>0.815</u> |
| | | Concordance index | <u>0.902</u> | <u>0.902</u> |
| | | AUROC (H vs. ML) | <u>0.914</u> | <u>0.914</u> |
| | | AUROC (HM vs. L) | <u>0.994</u> | <u>0.994</u> |

**Table 4. The control analysis on the stem cell dataset and wedge dataset measured by drugs.** Sotalol was selected as the example control drug in both datasets. The higher values between with and without control are underscored.

**Supplementary**

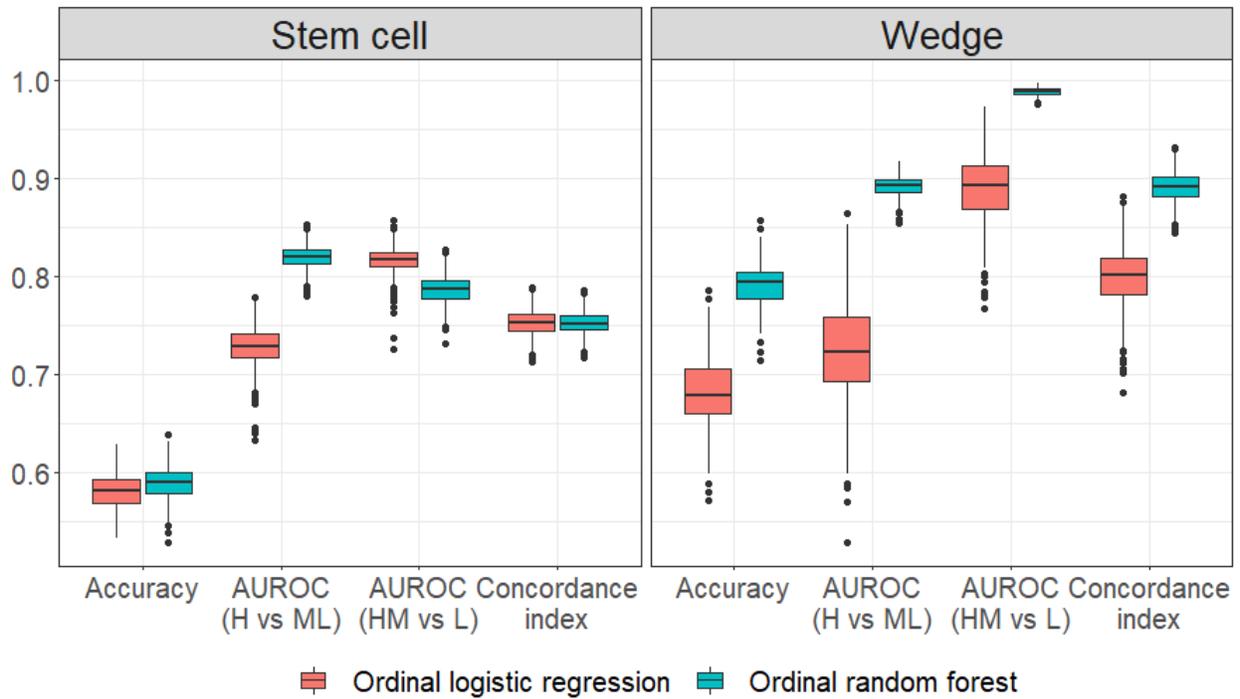

**Figure S1. The empirical distributions of model performance under stratified bootstrap measured by observations.** Four performance measurements of two models were calculated on the stem cell and wedge datasets.

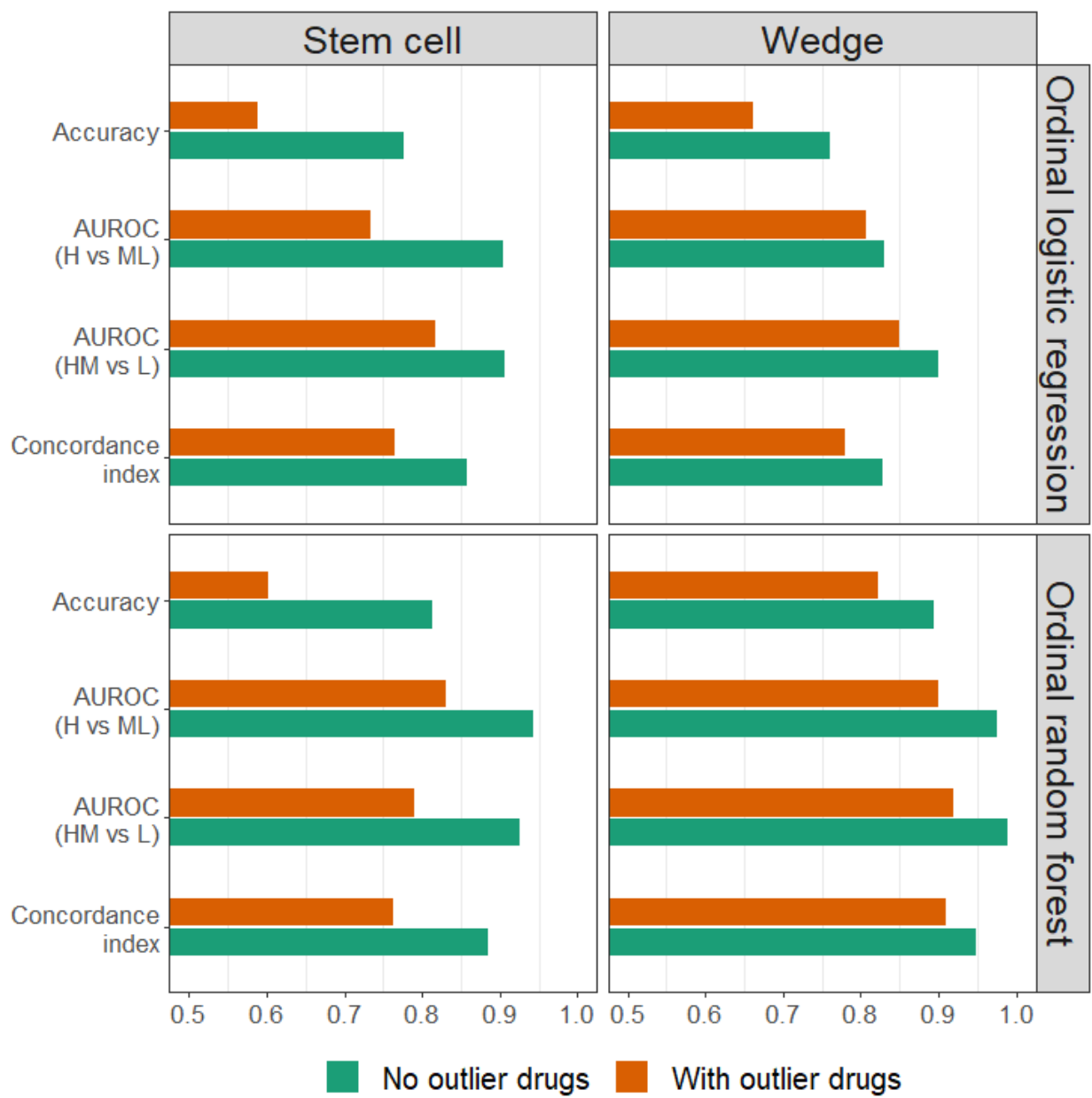

**Figure S2. The model performance before and after the removal of potential outlier drugs measured by observations.** The four measurements were calculated under four conditions: 1) ordinal logistic regression on the stem cell dataset; 2) ordinal logistic regression on the wedge dataset; 3) ordinal random forest on the stem cell dataset; 4) ordinal random forest on the wedge dataset.

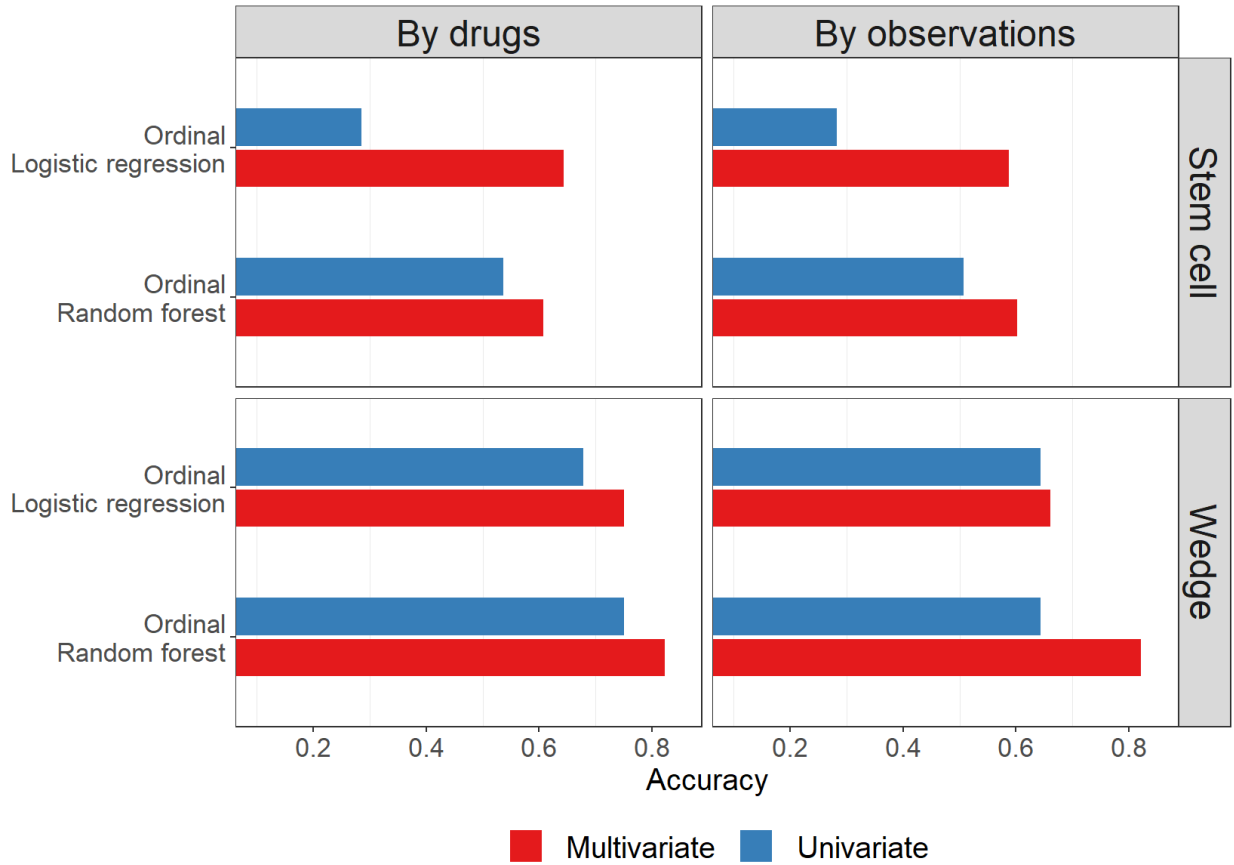

**Figure S3. The comparison of model prediction accuracy using one predictor (univariate) and all predictors (multivariate) in the dataset.** The single predictor in the univariate model is arrhythmia (whether drug-induced arrhythmias occurred at any concentration in any wells) in the stem cell dataset and c3v9 (interval between Q point and the end of T wave) in the wedge dataset.

| Low TdP Risk (9) | Intermediate TdP Risk (11) | High TdP Risk (8) |
|---|---|---|
| Diltiazem, Loratadine Ranolazine, Metoprolol Mexiletine, Nifedipine Nitrendpine, Tamoxifen Verapamil | Astemizole, Chlorpromazine Cisapride, Clarithromycin Clozapine, Domperidone, Droperidol, Ondansetron, Pimozide, Risperidone, Terfenadine | Azimilide, Bepridil Disopyramide, Dofetilide Ibutilide, Quinidine Sotalol, Vandetanib |

**Table S1. The 28 CiPA drugs in the stem cell dataset and wedge dataset.**

| Predictor | Description | Type |
|---|---|---|
| arrhythmia | Whether drug-induced arrhythmias occurred at any concentration in any wells | Binary |
| AMtwooutoffive | Whether drug-induced arrhythmias occurred at any concentration in more than 40% of wells | Binary |
| maxpro | Maximum repolarization change (ms) observed at any concentration | Continuous |
| fst_pro | Repolarization prolongation (ms) at the first drug concentration with statistically significant prolongation or shortening | Continuous |
| foldprolong | Drug concentration (folds over Cmax) at which the statistically significant (p-value $\leq 0.05$) repolarization prolongation was first observed | Continuous |
| foldaym | Drug concentration (folds over Cmax) when drug-induced arrhythmias were first observed | Continuous |
| pred7 | Drug-induced repolarization change (ms) at Cmax | Continuous |

**Table S2. The seven predictors in the stem cell dataset.**

| Predictor | Description | Type |
|---|---|---|
| c3v1 | TdP score proposed in the original study[14] | Continuous |
| c3v2 | Interval between the J point and the start of T wave (JT) | Continuous |
| c3v3 | Interval between the J point and the peak of T wave (JTP) | Continuous |
| c3v4 | Ratio of the JTP interval at concentration $\geq 0$ to concentration $= 0$ | Continuous |
| c3v6 | QS interval, measured at pace rate 2000 | Continuous |
| c3v7 | QS interval, measured at pace rate 500 | Continuous |
| c3v8 | Ratio of the QS interval at concentration $\geq 0$ to concentration $= 0$, measured at pace rate 500 | Continuous |
| c3v9 | Interval between Q point and the end of T wave (QTE) | Continuous |
| c3v10 | Ratio of the QTE interval at concentration $\geq 0$ to concentration $= 0$ | Continuous |
| c3v11 | Ratio of QT interval to QS interval | Continuous |
| c3v12 | Ratio of the interval between the peak and the end of T wave (TPE) to QT interval | Continuous |
| c3v13 | Ratio of TPE interval[52] to QT interval | Continuous |
| c3v14 | TPE Interval | Continuous |
| c3v15 | Ratio of TPE interval at concentration $\geq 0$ to concentration $= 0$ | Continuous |
| c3v16 | Score as a function of EAD[14] | Continuous |

**Table S3. The 15 predictors in the wedge dataset.** Except for extra notes, all predictors were obtained from an electrocardiogram at a 2000 pace rate.

| Dataset | Model | Accuracy (Three-category) | AUROC (High vs. Intermediate-or-Low) | AUROC (High-or-Intermediate vs. Low) | Concordance index (Three-category) |
|---|---|---|---|---|---|
| Stem cell | Ordinal logistic regression | 0.588 | 0.734 | <u>0.817</u> | <u>0.764</u> |
| Stem cell | Ordinal random forest | <u>0.603</u> | <u>0.840</u> | 0.791 | 0.762 |
| Wedge | Ordinal logistic regression | 0.661 | 0.808 | 0.849 | 0.780 |
| Wedge | Ordinal random forest | <u>0.822</u> | <u>0.900</u> | <u>0.919</u> | <u>0.909</u> |

**Table S4. The model performance measured by observations (point estimate).** Three-category prediction accuracy, two AUROCs, and concordance index are shown for ordinal logistic regression and ordinal random forest. The higher values between the two models are underscored.

|  |  | True Risk |  |  |
|---|---|---|---|---|
|  |  | Low | Intermediate | High |
| Prediction | Low | 6 | 3 | 0 |
|  | Intermediate | 3 | 7 | 3 |
|  | High | 0 | 1 | 5 |

**Table S5. The confusion matrix of drug risk prediction by ordinal logistic regression on the stem cell dataset.** Six drugs are underestimated, and four drugs are overestimated. The model shows aggressive predictive behavior.

|  |  | True Risk |  |  |
|---|---|---|---|---|
|  |  | Low | Intermediate | High |
| Prediction | Low | 5 | 2 | 0 |
|  | Intermediate | 4 | 7 | 3 |
|  | High | 0 | 2 | 5 |

**Table S6. The confusion matrix of drug risk prediction by ordinal random forest on the stem cell dataset.** Five drugs are underestimated, and six drugs are overestimated. The model shows conservative predictive behavior.

|              | True Risk |              |      |
|--------------|-----------|--------------|------|
| Prediction   | Low       | Intermediate | High |
| Low          | 8         | 1            | 1    |
| Intermediate | 1         | 8            | 2    |
| High         | 0         | 2            | 5    |

**Table S7. The confusion matrix of drug risk prediction by ordinal logistic regression on the wedge dataset.** Four drugs are underestimated, and three drugs are overestimated. The model shows aggressive predictive behavior.

|              | True Risk |              |      |
|--------------|-----------|--------------|------|
| Prediction   | Low       | Intermediate | High |
| Low          | 8         | 0            | 0    |
| Intermediate | 1         | 9            | 2    |
| High         | 0         | 2            | 6    |

**Table S8. The confusion matrix of drug risk prediction by ordinal random forest on the wedge dataset.** Two drugs are underestimated, and three drugs are overestimated. The model shows conservative predictive behavior.

| Measured by  | Ordinal logistic regression | Ordinal random forest | Multinomial logistic regression | Multinomial random forest | Classical ordinal logistic regression |
|--------------|-----------------------------|-----------------------|---------------------------------|---------------------------|---------------------------------------|
| Observations | 0.588                       | <u>0.603</u>          | 0.573                           | 0.592                     | 0.550                                 |
| Drugs        | <u>0.643</u>                | 0.607                 | <u>0.643</u>                    | 0.607                     | 0.607                                 |

**Table S9. The comparison of three-category prediction accuracy among different models on the stem cell dataset.** Ordinal logistic regression and ordinal random forest are described in the main text. Multinomial logistic regression and multinomial random forest do not utilize ordinal information in the drug TdP risk. The classical ordinal logistic regression is described in [62]. The higher accuracy among different models is underscored.

| Measured by | Ordinal logistic regression | Ordinal random forest | Multinomial logistic regression | Multinomial random forest | Classical ordinal logistic regression |
|---|---|---|---|---|---|
| Observations | 0.661 | <u>0.840</u> | 0.741 | 0.830 | 0.741 |
| Drugs | 0.785 | <u>0.822</u> | 0.750 | 0.785 | 0.750 |

**Table S10. The comparison of three-category prediction accuracy among different models on the wedge dataset.** Ordinal logistic regression and ordinal random forest are described in the main text. Multinomial logistic regression and multinomial random forest do not utilize ordinal information in the drug TdP risk. The classical ordinal logistic regression is described in [62]. The higher accuracy among different models is underscored.

| Dataset | Model | Measurement | Accuracy (Three-category) | AUROC (High vs. Intermediate-or-Low) | AUROC (High-or-Intermediate vs. Low) | Concordance index (Three-category) |
|---|---|---|---|---|---|---|
| Stem cell | Ordinal logistic regression | 95% CI | (0.548, 0.615) | (0.691, 0.764) | (0.793, 0.837) | (0.730, 0.775) |
| | | Mean | 0.581 | 0.729 | <u>0.817</u> | <u>0.753</u> |
| | Ordinal random forest | 95% CI | (0.560, 0.619) | (0.799, 0.841) | (0.761, 0.813) | (0.731, 0.774) |
| | | Mean | <u>0.590</u> | <u>0.820</u> | 0.787 | <u>0.753</u> |
| Wedge | Ordinal logistic regression | 95% CI | (0.608, 0.759) | (0.628, 0.815) | (0.828, 0.949) | (0.743, 0.853) |
| | | Mean | 0.683 | 0.725 | 0.890 | 0.800 |
| | Ordinal random forest | 95% CI | (0.741, 0.831) | (0.870, 0.911) | (0.981, 0.995) | (0.862, 0.917) |
| | | Mean | <u>0.790</u> | <u>0.892</u> | <u>0.988</u> | <u>0.891</u> |

**Table S11. The summary statistics of model performance under stratified bootstrap measured by observations.** The 95% confidence intervals and means of four measurements were calculated for ordinal logistic regression and ordinal random forest. The higher mean values between the two models are underscored.

| Dataset | Model | Measurement | Without control | With control |
|---|---|---|---|---|
| Stem cell | Ordinal logistic regression | Accuracy | 0.587 | <u>0.630</u> |
| | | Concordance index | 0.750 | <u>0.780</u> |
| | | AUROC (H vs. ML) | 0.689 | <u>0.732</u> |
| | | AUROC (HM vs. L) | 0.780 | <u>0.826</u> |
| | Ordinal random forest | Accuracy | 0.560 | <u>0.572</u> |
| | | Concordance index | 0.723 | <u>0.731</u> |
| | | AUROC (H vs. ML) | 0.797 | <u>0.830</u> |
| | | AUROC (HM vs. L) | 0.774 | <u>0.778</u> |
| Wedge | Ordinal logistic regression | Accuracy | 0.657 | <u>0.670</u> |
| | | Concordance index | 0.778 | <u>0.785</u> |
| | | AUROC (H vs. ML) | 0.741 | <u>0.797</u> |
| | | AUROC (HM vs. L) | <u>0.896</u> | 0.895 |
| | Ordinal random forest | Accuracy | <u>0.824</u> | <u>0.824</u> |
| | | Concordance index | <u>0.911</u> | <u>0.911</u> |
| | | AUROC (H vs. ML) | <u>0.895</u> | <u>0.895</u> |
| | | AUROC (HM vs. L) | <u>0.988</u> | <u>0.988</u> |

**Table S12. The control analysis on the stem cell dataset and wedge dataset measured by observations.** Sotalol was selected as the example control drug in both datasets. The higher values between with and without control are underscored.